\documentclass[review]{elsarticle}
\usepackage[utf8]{inputenc}
\usepackage{subfig}
\usepackage{multirow}
\usepackage{amsmath,amssymb,amsfonts}
\DeclareMathOperator*{\argmin}{argmin} 

\usepackage{graphicx}
\usepackage{textcomp}
\usepackage{soul}
\usepackage{adjustbox}
\usepackage{mathrsfs}

\usepackage[dvipsnames]{xcolor}

\usepackage{booktabs, tabularx, makecell, array}

\newcolumntype{L}[1]{>{\raggedright\arraybackslash}m{#1}}
\usepackage{siunitx}
\usepackage[ruled,vlined,linesnumbered]{algorithm2e}
\usepackage{algpseudocode}
\usepackage{pdflscape}
\usepackage{afterpage}
\usepackage{float}

\newtheorem{theorem}{Theorem}[section]

\newtheorem{definition}[theorem]{Definition}
\newtheorem{remark}{Remark}

\SetKwRepeat{Do}{do}{while}

\title{Optimal Load Balancing and Assessment of Existing Load Balancing Criteria}

\author[1]{Anthony Boulmier\corref{cor1}}
\author[2]{Nabil Abdennadher}
\author[1]{Bastien Chopard}

\address[1]{University of Geneva, Departement of Computer Science, Route de Drize 7, 1227 Carouge, Switzerland
}
\address[2]{University of Applied Sciences and Arts, Western Switzerland (HES-SO), Rue de la Prairie 4, 1202 Geneva, Switzerland}

\cortext[cor1]{Corresponding author\\Email address: anthony.boulmier@unige.ch}

\begin{document}
\begin{frontmatter}
\begin{abstract}
    Parallel iterative applications often suffer from load imbalance, one of the most critical performance degradation factors. Hence, load balancing techniques are used to distribute the workload evenly to maximize performance. A key challenge is to know \textit{when} to use load balancing techniques. In general, this is done through load balancing criteria, which trigger load balancing based on runtime application data and/or user-defined information. In the first part of this paper, we introduce a novel, automatic load balancing criterion derived from a simple mathematical model. In the second part, we propose a branch-and-bound algorithm to find the load balancing iterations that lead to the optimal application performance. This algorithm finds the optimal load balancing scenario in quadratic time while, to the best of our knowledge, this has never been addressed in less than an exponential time. Finally, we compare the performance of the scenarios produced by state-of-the-art load balancing criteria relative to the optimal load balancing scenario in synthetic benchmarks and parallel N-body simulations. In the synthetic benchmarks, we observe that the proposed criterion outperforms the other automatic criteria. In the numerical experiments, we show that our new criterion is, on average, $4.9\%$ faster than state-of-the-art load balancing criteria and can outperform them by up to $17.6\%$. Moreover, we see in the numerical study that the state-of-the-art automatic criteria are at worst $47.4\%$ slower than the optimum and at best $16.5\%$ slower. 
\end{abstract}


\end{frontmatter}

\section{Introduction}

Parallel iterative applications often exhibit an irregular computational scheme that may create load imbalance over time. Load imbalance is a major performance degradation factor. For that purpose, dynamic load balancing mechanisms are used throughout the application execution to keep processing elements' workloads evenly distributed and their communications minimized. Those mechanisms involve two separate questions \textit{how} and \textit{when} to load balance~\cite{Pearce2014}. ``How to load balance'' is related to finding the algorithm that divides the computational domain (partitioning algorithm) into several pieces that are distributed (mapping algorithm) on the available processing elements while minimizing their communications. ``When to load balance'' defines the particular iterations (i.e., a scenario) at which the load balancing mechanism (i.e., using the partitioning and mapping algorithm) is required. Their goal is to minimize the application wall time. 

``How to load balance'' has been explored by several authors over the years leading to various partitioning and mapping algorithms. In particular, the partitioning algorithm consists of solving a balancing graph partitioning problem, which is known to be NP-Complete~\cite{Garey1979}. Hence, heuristics have been developed, exhibiting good balancing capabilities for various types of problems. Among the most famous, recursive coordinate bisection (RCB)~\cite{Simon1997HowBisection}, space-filling curves (SFC)~\cite{Borrell2019ParallelBalancing}, recursive spectral bisection~\cite{VanDriessche1995AnBalancing}, and METIS (multilevel k-way)~\cite{Karypis1998AGraphs} can be mentioned. For more sophisticated techniques, we suggest the reader to refer to~\cite{Fleissner2008, Fattebert2012, Furuichi2017IterativeInteractions}. However, it is difficult for scientists to know how well a particular technique will perform on their own problem. Moreover, due to the complexity of modern algorithms and the lack of ``plug and play'' libraries, scientists often use the most famous load balancing techniques, which may not be optimal for their problem. In addition, we pointed out in a previous work that researchers should not select a load balancing technique only based on its capability to correct imbalance but also during how many iterations it keeps a low level of imbalance~\cite{Boulmier2019OnApplications}. This further increases the difficulty to select the most optimal technique. To overcome this challenge, researchers have proposed algorithms to select automatically the most suitable load balancing techniques based on application data~\cite{Pearce2012, Menon2016, Boulmier2017}. 

``When to load balance'' is a challenging problem that involves finding the iterations at which a parallel iterative application should trigger its load balancing mechanisms to maximize performance. Herein, we refer to the optimal load balancing scenario as the sequence of iterations where the load balancing is applied such that the application wall time is minimized. As the load balancing itself has a cost (compute the new partition and migrate the data), one (most of the time) can not simply re-balance every iteration because the load balancing cost $C$ overcomes the performance gain. To find the optimal scenario for analysis purposes, one straightforward way is to try every load balancing scenario and keep the one that yields the best performance. However, this is unfeasible in practice, even for a small number of iterations. Indeed, for an application comprising of $\gamma$ iterations, the number of scenarios is $2^\gamma$. 

In the literature, load balancing criteria have been proposed to decide whether the load balancing mechanism should be triggered or not. A load balancing criterion is a condition based on application information and/or user data. One of the most straightforward criteria re-balances the application every $T$ iterations enabling the correction of recurring imbalance. However, this is inefficient when load imbalance exhibits a non-periodic pattern. Some more sophisticated criteria use mathematical models taking into account collected data, such as the unbalancing pace (i.e., workload increase rate), the load balancing cost, the expected scalability, the maximum authorized imbalance, and others. For instance, Marquez et al.~\cite{MarquezClaudio2013AApplications} propose to apply the load balancing algorithm when at least one of the processing elements is below (respectively above) a pre-defined workload lower bound (respectively upper bound). Procassini et al.~\cite{Procassini2004LoadCalculations} predict the time per iteration post load balancing using an estimation of the efficiency's improvement and trigger the re-balancing mechanism when the increase in time per iteration is greater than the load balancing cost. Menon et al.~\cite{H.MenonandN.JainandG.ZhengandL.Kale2012} propose to re-balance the application when the cumulative load imbalance (i.e., the sum of the current imbalance over time) overcomes the load balancing cost. Pearce~et~al.~\cite{Pearce2012} perform a cost-benefit analysis of the load balancing process. They use a load model to estimate both the cost of load balancing with various algorithms and the benefit of correcting the imbalance. They activate the load balancing mechanism if its benefit is greater than its cost. Finally, the wide choice of criteria makes the choice of a suitable criterion difficult due to the lack of rigorous comparative studies. Worse, because it is hard to find the optimal load balancing scenario among all the possible candidates, there is no clue how far the performance of the scenario produced by a load balancing criterion is from the optimal scenario's performance. Therefore, finding the optimal scenario and quantifying its performance is an important and challenging task. 

This paper introduces a load balanced application theoretical model to derive a novel, automated load balancing criterion that performs at worse on par with state-of-the-art load balancing criteria. Moreover, we propose a new method derived from the $A{^*}$ algorithm~\cite{Hart1968APaths} to find the iterations at which the load balancing mechanism must be used to obtain optimal performances. This method can be applied to real applications and synthetic benchmarks built with our theoretical model. Then, we use the optimal scenario to evaluate the performance of several state-of-the-art load balancing criteria on various synthetic benchmarks. 
Such a study provides insights into the performance gap between state-of-the-art criteria and the optimum. Finally, we implement our novel algorithm in an N-body simulation and discuss the difference of performance and behavior between state-of-the-art load balancing criteria and the optimal scenario. The result of our efforts also includes two implementations of our novel algorithm. The first is a standalone package for studying optimal scenarios within synthetic benchmarks, and the second is an implementation for real applications. 

Section~\ref{section:problem_statement} proposes a definition of the load balancing decision problem and introduces the challenges to solve it.
Section~\ref{section:background_work} presents the background works related to load balancing criteria.
Section~\ref{sec:LBAppModel} introduces a model for parallel applications with dynamic load balancing and shows how to derive a novel, fully automatic load balancing criterion based on the past and current behavior.
Section~\ref{sec:OptimalScenarioAlgorithm} presents an efficient algorithm to find the optimal load balancing scenario.  
Section~\ref{sec:comparison} assesses the performance of load balancing criteria with respect to the optimal scenario in synthetic benchmarks and within a parallel N-body simulation.
Section~\ref{sec:conclusion} concludes this work and proposes insight for future works.

\section{The Load Balancing Decision Problem}
\label{section:problem_statement}

Consider an iterative parallel application (e.g., N-body, computational fluid dynamics, etc.) comprising of $\gamma$ iterations. Computing such an application in parallel on $P$ processing elements requires distributing its workload among the processing units used for the computation while minimizing communications. The time per time-step is equal to the time of the slowest (or most loaded) processing element due to synchronization mechanisms at the end of each iteration. To maximize efficiency, the workload attributed to each processing element must be roughly equal at each iteration. This is achieved through load balancing algorithms that mitigate the load imbalance penalty. For parallel applications that do not exhibit a dynamic nature, only one load balancing is required at the beginning of its execution. This is usually called static load balancing. In contrast, when the processing elements' workload is not the same from iteration to iteration, several load balancing steps may be required. This is known as dynamic load balancing. We call the set of iterations at which the load balancing algorithm is used the ``load balancing scenario''. The processing elements must coordinate and take a load balancing decision at each iteration (re-balancing or not) to create a scenario. This decision process leads to $2^\gamma$ possible scenarios where $\gamma$ is the number of iterations. The dynamic load balancing decision problem consists of finding the optimal scenario, minimizing the application wall time.

\begin{definition}[Dynamic Load Balancing Decision Problem]
Given an application comprising of $\gamma$ iterations and $P$ processing elements, find the set of iterations $\sigma^*$ (i.e., the scenario) at which the load balancing mechanism must be activated such that the application wall time is minimized. 
\end{definition}
This decision problem is an optimization problem in which we look for
\begin{equation}\label{eq:opt_problem}
    \sigma^* = \argmin\limits_{\sigma \in \mathcal{S}} \ T(\sigma),
\end{equation} where $T(\sigma)$ is a function returning the application wall time given a load balancing scenario and $\sigma$ is a particular scenario among the $2^\gamma$ possible ones ($S$). Note that $T(\sigma)$ can either be modeled by an equation or computed by the application code itself (i.e., actually measured on a computer).

Solving this problem is non-trivial as the load balancing benefit usually depends on the application's future behavior, the moment at which the load balancing is applied, and the success of the data partitioning. Let us imagine an application where the load imbalance is ephemeral. Molecular dynamic applications may see such behaviors. For instance, the particle density across the computational domain can change periodically due to some forces. Therein, it is unclear whether re-distributing the particles would be beneficial due to the load balancing cost. To accurately answer this question, one would have to predict that such behavior happens. This implies that the application is predictable (in the long term), which appears to be unfeasible~\cite{Pearce2012}. Therefore, load balancing decisions can only be built on strong and reliable metrics based on prior data.

Asking whether re-balancing the workload is required or not depends on multiple factors. In the literature, scientists base their decision on load balancing criteria that employ various metrics such as the parallel efficiency, the load imbalance, the iteration index, the min/max workload, and others. For a review of load imbalance metrics, we suggest the reader to refer to~\cite{Rodrigues2016StudyApplications}. We define the criteria that use local information as \textit{local criteria}, whereas the other criteria are considered as \textit{global criteria}. A local information is a data that is related to a single processing element, such as the current processing element workload, the processing element workload increase rate, etc. In contrast, a global information concerns all processing elements, such as the time per iteration, the average workload, or the load balancing cost. In the next section, we dig into more details in the various load balancing criteria proposed over time by researchers.

\section{Background works}
\label{section:background_work}

In the literature, scientists often use straightforward load balancing criteria while it is well known that without fine-tuning, they provide poor performance~\cite{Pearce2012, H.MenonandN.JainandG.ZhengandL.Kale2012}. For instance, Fattebert~et~al.~\cite{Fattebert2012}, Offenhäuser~\cite{Offenhauser2017Load-balanceSystems}, and Lieber~et~al.~\cite{Lieber2018HighlyModeling} chose to load balance their application respectively every $100$, $1000$, and $180$ iterations while Ishiyama~et~al.~\cite{Ishiyama2012} re-balance every iteration. The rationales behind these choices are manifold. Some argue that the load balancing cost is negligible compared to load imbalance~\cite{Ishiyama2012}, while others use application knowledge to tune their criterion. However, we will see later in this paper that a bad load balancing criterion can suffer from a huge performance penalty compared to the optimal scenario. Indeed, it is likely that many of the load balancing calls are unnecessary, ill-timed, or worse, the application may still suffer from load imbalance. For that purpose, researchers have tried to develop more sophisticated and generic criteria that provide better overall performance. 

Marquez et al.~\cite{MarquezClaudio2013AApplications} have proposed a load balancing criterion based on an acceptable workload variation range for agent-based simulations. The idea is to trigger the load balancing mechanism if any agent's workload goes outside of a comfort zone defined by a minimal acceptable workload $W_{\text{min}}$ and a maximal acceptable workload $W_{\text{max}}$. In other words, when the following condition is true:
\begin{equation}
    W_{p} < W_{\text{min}} \  \text{or} \ W_{\text{max}} < W_{p} \ \exists p=1..P.
    \label{eq:marquez1}
\end{equation}
This criterion is considered local as the formula uses the local workload of processing element $W_p$. The formula proposed by Marquez~et~al. can be implemented using a ``tolerance factor'' $\xi$, which specifies how far a single processing element can get away from the average workload. Then, Equation~\ref{eq:marquez1} can be rewritten 
\begin{equation}
    W_{p} < \frac{(1-\xi)}{P}  \sum_1^P W_{p}\ \text{or} \ \frac{(1+\xi)}{P} \sum_1^P W_{p} < W_{p}.
\end{equation}
Indeed, the tolerance factor has to be tuned by hand as the value may differ from application to application, making it difficult to find a good value for this parameter. Moreover, within a single application, the tolerance factor that provides the best performance may change over time. Unfortunately, an automatic selection of the acceptable workload range has never been proposed. 

Procassini et al.~\cite{Procassini2004LoadCalculations} use a different strategy to automatically load balance HPC applications. Their global criterion redistributes the workload whether the performance improvement due to load balancing plus the load balancing cost is greater than a fraction of the current time per iteration. In other words, the load balancing mechanism is triggered at iteration $t$ when the following condition is true: 
\begin{equation}\label{eq:proca}
     T_{\text{withLB}}(t) + C < \rho * T_{\text{withoutLB}}(t),
\end{equation} where $T_{\text{withLB}}(\cdot)$ is the iteration time after load balancing, $C$ is the load balancing cost in seconds, $\rho$ is the desired increase in performance post load balancing, and $T_{\text{withoutLB}}(\cdot)$ is the iteration time before load balancing. In their paper, they used $\rho=0.9$. However, the same idea can be generalized for any $\rho \in \mathbb{R}^{>0}$. Procassini~et~al. estimate the time per iteration post load balancing by decreasing the current time per iteration proportionally to the expected increase in performance due to load balancing. This reads
\begin{equation}
    T_{\text{withLB}}(t) = \frac{\varepsilon_{\text{pre}}(t)}{\varepsilon_{\text{post}}(t)} * T_{\text{withoutLB}}(t),
\end{equation} where $\varepsilon_{\text{post}}$ (resp. $\varepsilon_{\text{pre}}$) is the parallel efficiency post (resp. pre) load balancing step. While the parallel efficiency post load balancing has to be estimated based on prior data, the efficiency before load balancing is computed with 
\begin{equation*}
    \varepsilon_{\text{pre}}(t) = \frac{T_{\text{seq}}(t)}{P*T_{\text{par}}(t)}.
\end{equation*} The presence of the factor $\rho$, which must be fixed by hand, makes the tuning of this criterion for optimal performance difficult. Note that Lieber~et~al.~\cite{Lieber2018HighlyModeling} also implemented an ``auto-mode'' into their application (FD4), which employs a simple cost-benefit analysis of the load balancing process. The criterion utilized therein is analog to Equation~\ref{eq:proca}, except that they use $\rho = 1$ and they estimates the time post-load balancing using data collected from previous load balancing steps.

Menon et al.~\cite{H.MenonandN.JainandG.ZhengandL.Kale2012} have shown that the optimal load balancing scenario for a parallel iterative application, where the maximum and average load can be modeled linearly with time, is a fixed re-balancing frequency. The load balancing time interval $\tau$ is equal to the amount of iteration required by the cumulative load imbalance to reach the load balancing cost $C$. When the workload increase rate is constant, it can be computed by the following formula:
\begin{equation}\label{eq:menon_ref}
    \tau = \sqrt{\frac{2C}{\alpha}}, 
\end{equation}
where $C$ is the load balancing cost in seconds and $\alpha$ is the difference in the time-per-iteration increase rate between the ``slowest'' processing element and the average time-per-iteration increase rate. They derived this global criterion by minimizing the time with respect to the load balancing time interval. Like the criterion proposed by Procassini et al.~\cite{Procassini2004LoadCalculations}, the information used therein is measured and updated throughout application execution. For instance, the load balancing cost $C$ has to be estimated while the maximum and average workload increase rates are measured at runtime. We refer to criteria analog to Menon criterion as Menon's like criteria. 

Recently, Zhai et al.~\cite{Zhai2018} have used Menon criterion to improve the performance of CMT‐nek. CMT-Nek is a compressible multiphase turbulance application, which enhance the physics of the CESAR Nek5000 application. In particular, they proposed to compute the cumulative time-per-iteration degradation $\mathcal{D}$ during application execution and to trigger a load balancing call when it has reached the load balancing cost $C$, as suggested by Menon, leading to this global load balancing criterion: 
\begin{equation*}
    \mathcal{D} \geq C,
\end{equation*}
where the cumulative time-per-iteration degradation $\mathcal{D}$ from the last load balancing iteration LB$_p$ up to the current iteration $t$ is computed using
\begin{equation*}
    \mathcal{D} = \sum_{i=LB_p}^{\text{t}} \Big( T_{\text{median}}(i, i-2) - T_{\text{avg}}(\mathbb{P}) \Big),
\end{equation*} where $T_{\text{avg}}(\mathbb{P})$ is the average time per time-step over an evaluation phase $\mathbb{P}$ and $T_{\text{median}}(i, i-2)$ is the median time per time-step among the three last iterations.

One difficulty for HPC developers regarding load balancing is to choose the good load balancing criteria. Indeed, all the criteria available in the literature bring confusion and only a few are backed up by a rigorous theory. We summarize the load balancing criteria described above in Table~\ref{tab:criteria} to ease the choice of HPC researchers. This table details what we find to be the most useful properties of load balancing criteria. In addition, in Section~\ref{sec:OptimalScenarioAlgorithm}, we propose an efficient branch-and-bound algorithm for finding the optimal load balancing scenario to compare load balancing criteria relative to the optimum and help the selection of load balancing criteria.  

\vfill

\begin{table}[htp]
\centering
\setlength{\tabcolsep}{5pt}
\renewcommand{\arraystretch}{2.5} 
\small
\makebox[\linewidth]{
\begin{tabularx}{1.4\textwidth}{L{2cm}L{3cm}L{3.8cm}L{1.5cm}L{2cm}L{3.5cm}}
\toprule
\toprule
 \thead[l]{Name}  &   
{\thead[l]{User defined \\ parameters}} &   
{\thead[l]{Use dynamic \\ information}} &   
{\thead[l]{Type}} &   
{\thead[l]{Foundation}} &   
{\thead[l]{Developed for}} \\
\midrule
Periodic & Load balancing period & - & Global & - & Any simulations \\

Marquez et al.~\cite{MarquezClaudio2013AApplications} & Tolerance factor $\xi$ & \shortstack{\textbullet PEs workload} & Local & Experiments & \shortstack[l]{Agent-based \\ simulations } \\
Procassini et al.~\cite{Procassini2004LoadCalculations} & Desired performance improvement post load balancing $\rho$  & \shortstack[l]{\textbullet Efficiency \\ \textbullet Load Balancing Cost} & Global & Experiments & \shortstack[l]{Monte-Carlo \\ Transport} \\
Menon et al.~\cite{H.MenonandN.JainandG.ZhengandL.Kale2012} & - & \shortstack[l]{\textbullet Imbalance Increase Rate \\ \textbullet Load Balancing Cost} & Global & Theory & Any simulations \\
Zhai et al.~\cite{Zhai2018} & Evaluation phase $\mathcal{P}$ & \shortstack[l]{\textbullet Imbalance Increase Rate \\ \textbullet Load Balancing Cost} & Global & Experiments & Any simulations \\
Our criterion & - & \shortstack[l]{\textbullet Imbalance Increase Rate \\ \textbullet Load Balancing Cost} & Global & Theory & Any simulations \\
\bottomrule
\end{tabularx}
}
\caption {Summary of available load balancing criteria. Note that the periodic criterion belongs to the common knowledge of load balancing. Tracking back its origin appears to be complicated.}
\label{tab:criteria}
\end{table}

\section{Modeling the Parallel Time of Load Balanced Applications}
\label{sec:LBAppModel}

To study the performance of the scenarios produced by load balancing criteria relative to the optimal load balancing scenario's performance, we propose a mathematical framework for computing the CPU time of load balanced parallel applications inspired by Menon's work~\cite{H.MenonandN.JainandG.ZhengandL.Kale2012}. First, let us consider a parallel execution on $P$ processors characterized by two functions, $\mu(t)$ and $m(t)$. The function $\mu(t)$ gives at each iteration $t$ the average load (i.e., the total load on the $P$ processors divided by $P$), whereas $m(t)$ gives the load of the slowest (or most loaded) processor at iteration $t$. Note that here, $\mu(t)$ and $m(t)$ are expressed with units of time. Furthermore, let assume that executing the load balancing mechanism always leads to perfect load balancing.

Second, let us define $T_{\text{par}}$ the parallel time on $\gamma$ iterations given by
\begin{equation} \label{eq:tpar_1}
    T_{\text{par}} = \int_0^\gamma m(t) \ \text{d}t = \int_0^\gamma m(t) - \mu(t) \ \text{d}t + \int_0^\gamma \mu(t) \ \text{d}t.
\end{equation} Note that this equation has been greatly inspired from the model of Menon~et~al.~\cite{H.MenonandN.JainandG.ZhengandL.Kale2012}, however, herein we relax the assumption that $m(t)$ and $\mu(t)$ are represented by line equations. Let us now divide the interval $[0,\gamma]$ in $n$ pieces $[s_i, s_{i+1}]$ with $s_0 = 0$, $s_{i+1} > s_i$ and $s_n = \gamma$. Moreover, we assume that load balancing steps are performed at iterations $s_i$ for $i=0,1,...n-1$ and they take an additional time $C$. Hence, Equation~\ref{eq:tpar_1} becomes
\begin{equation} \label{eq:tpar_2}
    T_{\text{par}} = \sum_{i=0}^{n-1} \big( \int_{s_i}^{s_{i+1}} u_i^*(t) \text{d}t + C \big) + \int_0^\gamma \mu(t) \ \text{d}t,
\end{equation} where $u_i^*(t)$ is the imbalance time metric proposed by DeRose~et~al.~\cite{DeRose2007DetectingSystems} defined as
\begin{equation*}
    u_i^*(t) = m(t) - \mu(t) \ \text{for} \ t \in [s_i, s_{i+1}].
\end{equation*} Also, we point out that load balancing is done at $s_0$ but not at the end of the execution (i.e., at $s_n$). Obviously, $m(t)$ resets to $\mu(t)$ after every load balancing step if the load is perfectly balanced. Thus, always $u_i^*(s_i) = 0$. To increase the readability of Equation~\ref{eq:tpar_2}, let us express it with the following change of variables
\begin{equation*}
    \tau_i = s_{i+1} - s_i \qquad u_i(x) = u_i^*(t - s_i).
\end{equation*} Equation~\ref{eq:tpar_2} now becomes
\begin{equation} \label{eq:tpar_3}
    T_{\text{par}} = \sum_{i=0}^{n-1} \big( \int_{0}^{\tau_i} u_i(x) \text{d}x + C \big) + \int_0^\gamma \mu(t) \ \text{d}t.
\end{equation} 

\begin{remark}
In general, $u_i(x)$ is unpredictable because it depends on $s_i$ as well as the load balancing technique itself. Indeed, the domain decomposition used by the load balancing mechanism affects the load imbalance growth. In section~\ref{sec:criteria-study:theory}, we give a possible solution to this challenge when we use our model as a framework for synthetic benchmarks.
\end{remark}

\paragraph{Derivation of Menon criterion} 
To derivate the criterion from Menon~et~al.~\cite{H.MenonandN.JainandG.ZhengandL.Kale2012} using Equation~\ref{eq:tpar_3} we need to set $u_i(x)$ as a linear equation such as
\begin{equation*}
    u_i(x) = u(x) = \alpha x.
\end{equation*} Then, Equation~\ref{eq:tpar_3} reads
\begin{equation*}
    T_{\text{par}} = \frac{\gamma}{\tau} \big( \int_{0}^{\tau} \alpha x \text{d}x + C \big) + \int_0^\gamma \mu(t) \ \text{d}t.
\end{equation*} Note that we obtain here the same equation as in the paper of Menon~et~al~\cite{H.MenonandN.JainandG.ZhengandL.Kale2012}. The optimal value of $\tau$ is then obtained by solving and isolating $\tau$ in
\begin{equation*}
    \frac{\partial T_{\text{par}}}{\partial \tau} = 0.
\end{equation*} That is, 
\begin{equation*}
\begin{split}
        \frac{\partial T_{\text{par}}}{\partial \tau} &= 0 \\
        -\frac{\gamma}{\tau^2} \big(\frac{\alpha\tau^2}{2} + C\big) + \frac{\gamma}{\tau}\alpha\tau&= 0\\
        \frac{\alpha}{2} -\frac{C}{\tau^2} &= 0 \\
        \tau &= \sqrt{\frac{2C}{\alpha}}
\end{split}
\end{equation*} 
It is worth noticing that for this value of $\tau$, one has
\begin{equation}\label{eq:MenonCrit}
    \int_0^\tau u(t) \text{d}t = \frac{\alpha \tau^2}{2} = C.
\end{equation} In other words, the load balancing mechanism must be used when the load imbalance metric $u(t) = m(t) - \mu(t)$ accumulated over the iterations reaches the load balancing cost $C$. For the sake of simplicity, this quantity reads
\begin{equation}\label{eq:U}
    U = \int_0^\tau u(t) \text{d}t.
\end{equation}

\begin{remark}\label{rem:proca=menon=ours}
In this case, where $u_i(x) = u(x) = \alpha x$, it is possible to obtain the optimal value of $\rho$ for Procassini criterion using Equation~\ref{eq:proca}. If  $\tau$ is the optimal load balancing interval when $u(x)$ is a linear equation, and the load balancing is perfect, the optimal value $\rho_\tau$ is \begin{equation}\label{eq:rhotau}
    \rho_\tau = \frac{T_{\text{withLB}}(\tau) + C}{T_{\text{withoutLB}}(\tau)} = \frac{\mu(\tau) + C}{\mu(\tau) + u(\tau)}.
\end{equation} Therefore, Procassini criterion is equal to Menon criterion provided that $u_i(x) = u(x) = \alpha x$ and that $\rho_\tau$ is employed. More generally, this indicates that for each load imbalance function $u(\cdot)$ there exists an optimal $\rho$ value. Unfortunately, as $u(\cdot)$ is in general unpredictable, computing $\rho_\tau$ seems highly challenging in practice.
\end{remark}

\begin{figure}
    \centering
    \includegraphics[width=\textwidth]{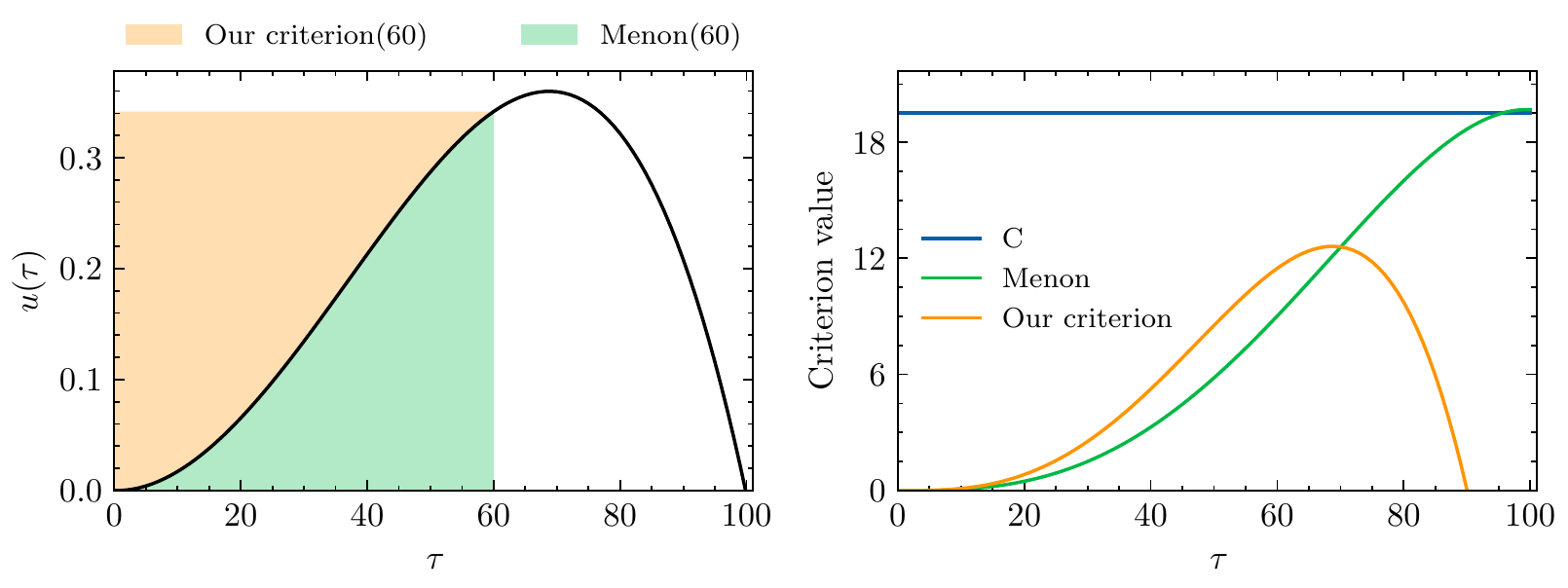}
    \caption{Toy example illustrating the key difference between our criterion and Menon criterion. The left figure shows a load imbalance that correct itself after a hundred iterations. The colored area in the left figure shows the criterion value of both criteria at iteration $60$. The right figure illustrates the evolution of the criterion value over the iterations. A key observation is that Menon criterion will apply the load balancing mechanism at iteration $96$ even though it is not needed, whereas our criterion successfully detects this situation. }
    \label{fig:our_criterion_vs_menon}
\end{figure}

\paragraph{Generalization for any $u(t)$} It is now possible to reformulate this result without assuming any particular form of $u(t)$. Starting from Equation~\ref{eq:tpar_3}, which now reads 
\begin{equation} \label{eq:tpar_4}
    T_{\text{par}} = \sum_{i=0}^{n-1} \big( \int_{0}^{\tau} u(x) \text{d}x + C \big) + \int_0^\gamma \mu(t) \ \text{d}t.
\end{equation} We obtain the optimal value of $\tau$ using the same methodology, which is solving and isolating $\tau$ in 
\begin{equation*}
    \frac{\partial T_{\text{par}}}{\partial \tau} = -\frac{\gamma}{\tau^2} \big(\int_0^\tau u(x)\text{d}x + C\big) + \frac{\gamma}{\tau}\alpha\tau = 0.
\end{equation*} The solution of this equation is
\begin{equation}
    \tau u(\tau) - \int_0^\tau u(x)\text{d}x = C,
\end{equation} which leads to a new global load balancing criterion that does not make any assumption on the function $u(t)$ that describes the load balancing metric over the iterations. This result differs from the one presented by Menon~et~al~\cite{H.MenonandN.JainandG.ZhengandL.Kale2012} because now a load balancing step is done when the area \textit{above} the load imbalance curve equals the load balancing cost $C$. To illustrate this, we propose in Figure~\ref{fig:our_criterion_vs_menon} a toy example showing the main difference between our criterion and Menon criterion as well as a plot of their value over time. In this example, the load imbalance is ephemeral starting at iteration $0$ and it grows until iteration $69$, then, it decreases until it reaches $u(100) = 0$. In the figure placed on the right, we see that Menon criterion applies a load balancing at iteration $96$ even though the load imbalance is almost completely corrected at this point. In contrast, we observe that our criterion is able to detect that such a situation does not need load balancing. Finally, the colored area in the left figure shows the criterion value of both criteria at iteration $60$. We see that, unlike Menon criterion, our criterion corresponds to the area between the load imbalance curve and $u(\tau)$. 

To have a better idea of the performance improvement, we might gain by using this criterion, we propose, in Section~\ref{sec:comparison}, a comparative study of the criteria presented in Section~\ref{section:background_work} on synthetic benchmarks and real N-body simulations. In the next section, we present an efficient algorithm for finding the optimal load balancing scenario, which we will use to rank the load balancing criteria as a function of their relative performance compared to the optimum.

\begin{remark}
Following the development of our theoretical model, it clearly appears that the exact solution of this problem can only be obtained using an exhaustive search. Indeed, we observed that this problem is recursive as the load balancing time intervals $s_i,s_{i+1}$ (i.e., the solution) are a part of the input data. This seems to prevent us from finding an analytical solution.
\end{remark}

\section{Finding the Optimal Load Balancing Scenario}
\label{sec:OptimalScenarioAlgorithm}
To analyze the performance of the scenario produced by load balancing criteria, it is essential to know the performance of the optimal load balancing scenario. To find this optimum, we need an efficient way to look for the optimal scenario among all the possible ones. Unfortunately, the number of possible scenarios grows exponentially with the number of iterations to compute (i.e.,~$\gamma$). For that reason, it is impossible to use brute force algorithms even for a small number of iterations. 

\begin{figure}[t]
    \centering
    \includegraphics[width=7cm]{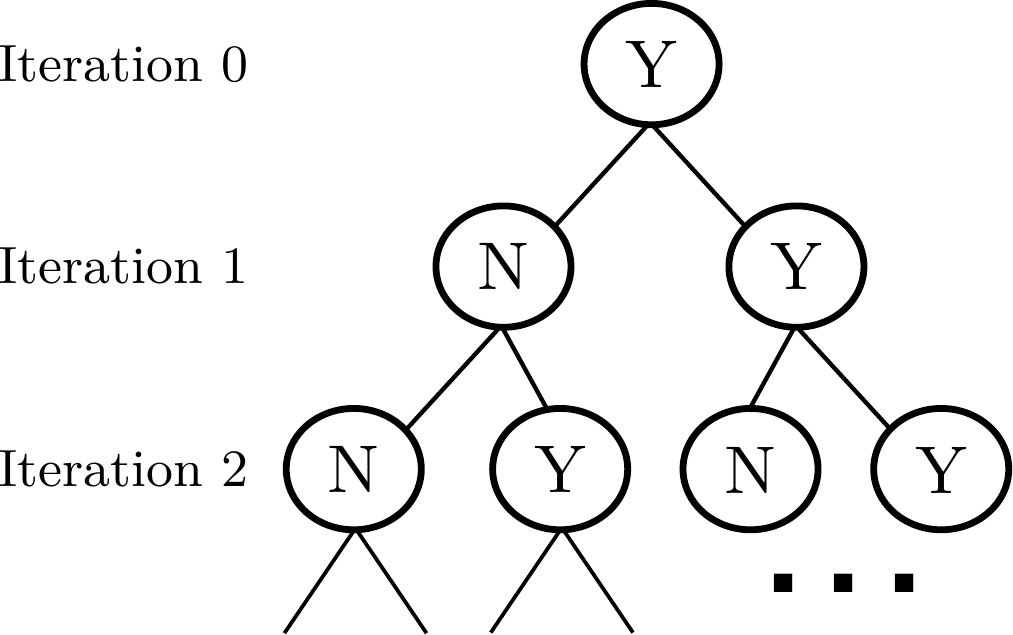}
    \caption{The load balancing decision problem organized as a binary tree. ``Y'' (respectively ``N'') means that the application is balanced (respectively not balanced) at the given iteration. In other words, left edges apply load balancing while right edges do not.}
    \label{fig:lbtree}
\end{figure}

To overcome this problem, we can organize the scenarios in a tree to use efficient tree search algorithms. Indeed, the load balancing decision problem fits well in a binary tree because a decision (using or not the load balancing mechanism) must be made at each iteration. The vertices represent the state of the application (balanced or not). The edges $e$ represent the process for going from an iteration to another (i.e., computing the iteration and applying, or not, the load balancing algorithm). The edge cost $\mathscr{C}(e)$ represents the CPU time for going from an iteration to the next. Figure~\ref{fig:lbtree} shows how load balancing decisions are organized as a binary tree. 

A load balancing scenario is defined as a path from the root (iteration $0$) to a leaf node (iteration $\gamma$). The cost of a path $p$ from the root node to any subsequent node, $\mathscr{C}(p)$, is the sum of the edge costs that belong to the path, which reads
\begin{equation*}\label{eq:cumulative_edge_cost}
    \mathscr{C}(p) = \sum_{e \in p} \mathscr{C}(e).
\end{equation*} As mentioned in Equation~\ref{eq:opt_problem}, the optimal scenario is the one that minimizes the path cost among all scenarios, minimizing the application wall time. 

\subsection{Load Balancing Tree Pruning} \label{sec:OptimalScenarioAlgorithm:TreePruning}
To reduce the tree size and the complexity of the search, we propose two steps: (i) to merge redundant load balancing nodes and (ii) to prune edges that belong to sub-optimal paths. We assume that the load balancing mechanism is independent of previous load balancing decisions in these two steps. This means that the workload of the processing units post load balancing does not depend on previous decisions but only on current information. Afterward, we apply the A$^*$ algorithm proposed by Hart~et~al.~\cite{Hart1968APaths}, in which we include these two optimizations, to find the optimal load balancing scenario. Note that the algorithm proposed herein has no practical uses in production, but is rather dedicated for analysis purposes because it requires some iterations to be executed multiple times.

\begin{figure}
    \centering
    \begin{minipage}[t]{0.48\textwidth}
        \centering
        \includegraphics[width=0.9\textwidth]{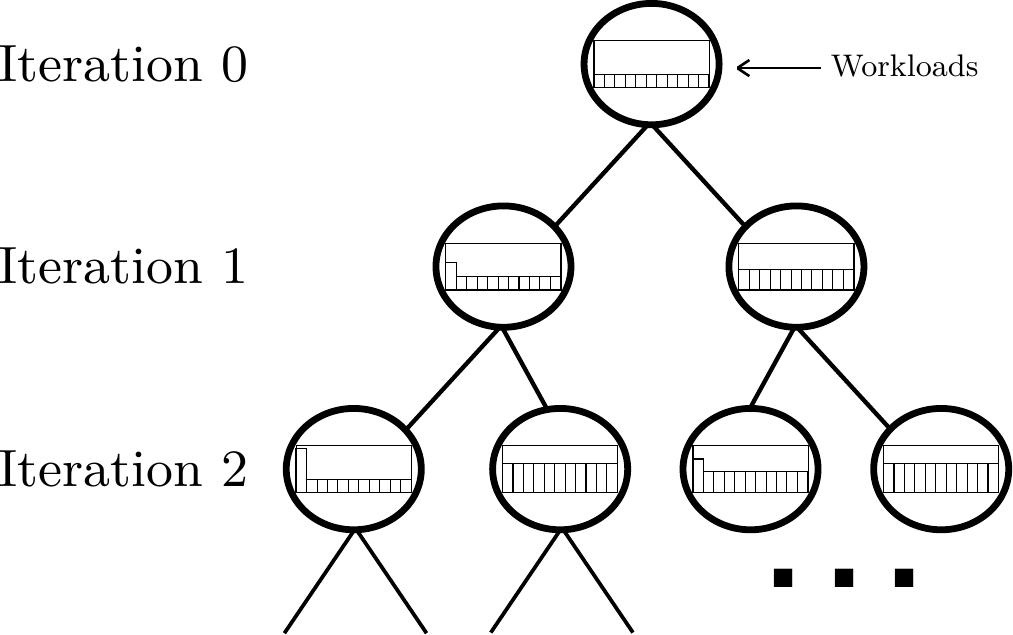} 
        \caption{The impact of load balancing on workloads within the load balancing tree before the merging process. The nodes that share the same data partitioning at the same iteration are redundant.}
        \label{fig:lb_tree_workloads}
    \end{minipage}\hfill
    \begin{minipage}[t]{0.48\textwidth}
        \centering
        \includegraphics[width=0.9\textwidth]{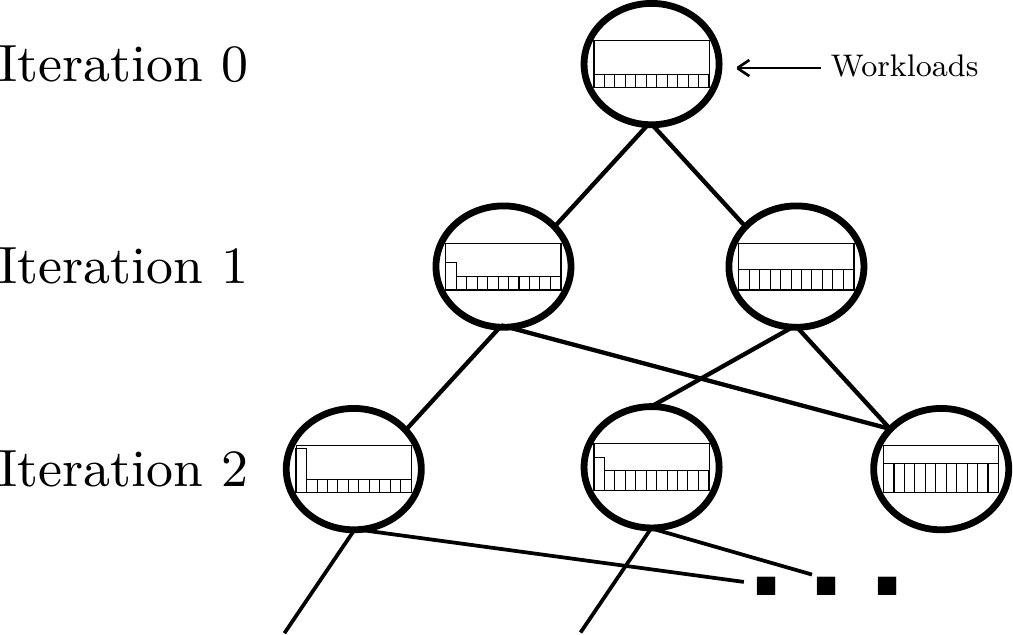} 
        \caption{The load balancing tree after the merging process, now, $i$ paths lead to a unique load balancing node (``Y'' node) at each iteration, where $i$ is the node's iteration.}
        \label{fig:lb_tree_workloads_merged}
    \end{minipage}
\end{figure}

\paragraph{Redundant nodes merging} 
As we saw previously in Section~\ref{sec:LBAppModel}, regardless of past decisions, the edge cost $\mathscr{C}(e)$ for going from iteration $i$ to the next is $C + \mu(i)$ if we performed a load balancing step. This is because the data partitioning after a load balancing call is independent of the previous decisions. This is illustrated in Figure~\ref{fig:lb_tree_workloads} which shows the processing elements' workload within the load balancing tree. Therein, we see that there is no difference between two load balancing nodes (flattened workloads) at the same level of the tree. The only thing that distinguishes these nodes is their cumulative cost. Therefore, load balancing nodes (i.e., ``Y'' nodes) at the same iteration can be merged.

\begin{figure}[t]
    \centering
    \includegraphics[width=10cm]{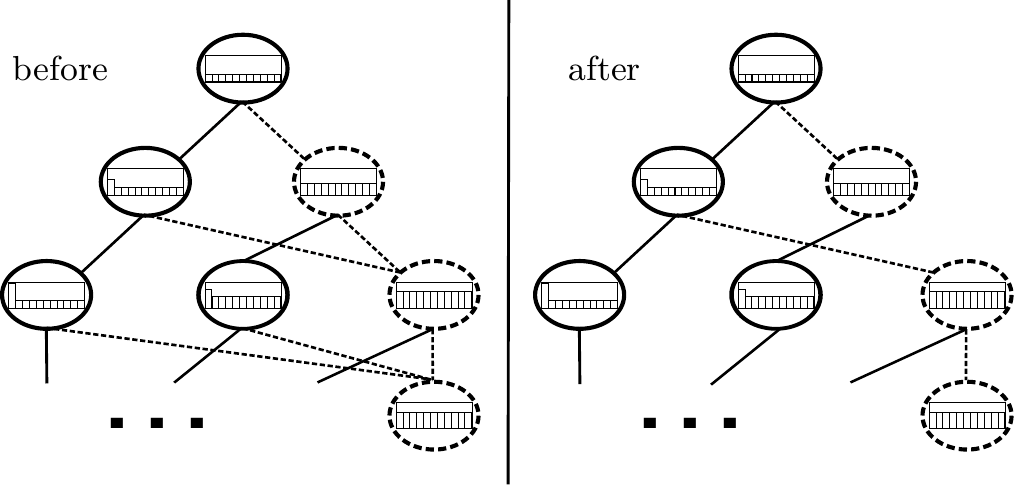}
    \caption{Example of the sub-optimal path elimination process. A dashed edge, from a dashed node (``Y'' node) to its parent, is removed if it does not belong to the shortest path from the root to the node. The total number of edges is reduced from exponential to linear in the number of iterations (i.e., the depth). }
    \label{fig:lb-tree-elimination}
\end{figure}

\paragraph{Sub-optimal path elimination} 
Merged nodes may have multiple paths leading to them, as illustrated in Figure~\ref{fig:lb_tree_workloads_merged}. The idea of sub-optimal path elimination is to find the shortest path from the root (iteration $0$) to each load balancing node (merged node) and to remove the other paths. 
Indeed, if a load balancing node $y$ is part of the final solution, then the shortest path from the root node to $y$ is also part of the solution. This is true if and only if the load balancing cost $C$ is independent of previous decisions, which is an assumption that we think to be reasonable. Finally, let assume that the load balancing node $y$ is a merged node at iteration $i$, therefore, $y$ has $i$ paths leading to it. Then, the shortest path $p_{0 \rightarrow y}^*$ is obtained by solving 
\begin{equation*} 
    p_{0 \rightarrow y}^* = \argmin\limits_{p_{0 \rightarrow y}^k \ \forall k = 1..i} \mathscr{C}(p_{0 \rightarrow y}^k),
\end{equation*} where $p_{0 \rightarrow y}^k$ is the kth path reaching node $y$. In practice, only the last edge of each sub-optimal path is removed because the previous edges belong to other paths. Figure~\ref{fig:lb-tree-elimination} illustrates a possible resulting tree after the sub-optimal path elimination process. 

Thanks to the pruning process, the size of the load balancing tree is drastically reduced. The number of vertices decreases from 
\begin{equation*}
    V = 2^{\gamma} - 1
\end{equation*}
to 
\begin{equation*}
    V = \sum_{i=0}^{\gamma-1} (i+1) = \frac{ \gamma(\gamma+1) }{ 2 }
\end{equation*}
 and the number of edges decreases from 
\begin{equation*}
    E = 2^{\gamma} - 2
\end{equation*}
to 
\begin{equation*}
    E = V - 1.
\end{equation*}
These two optimizations allow us to efficiently apply the A$^*$ algorithm to find the optimal load balancing scenario.

\subsection{Optimal Scenario Search Algorithm}

The $A^*$ algorithm~\cite{Hart1968APaths} is a well-known path search algorithm. It aims at finding the path from a source node to a destination with the smallest cost. Besides, $A^*$ is optimal and complete, which means that it will finish and it will find the solution if one exists. This is done by keeping a list of paths and extending those paths, one edge at a time until the destination is reached. At each iteration, $A^*$ extends the path that minimizes the cost equation
\begin{equation}
    f(n) = g(n) + h(n),
\end{equation} where $n$ is a candidate node, $g(n)$ is the total cost to reach that node, and $h(n)$ is an optimistic estimation of the path cost from $n$ to the destination node (i.e., the solution)~\cite{Hart1968APaths}. In our case, we model $g(n)$ as the time taken by the application to reach a particular iteration given a load balancing scenario. $h(n)$ represents the computation time from a particular iteration to the end of the application, given no load imbalance. This mathematically reads
\begin{equation*}
    h(n) = \sum_{j=i}^{\gamma} \mu(j),
\end{equation*} where $i$ corresponds to the node's iteration (i.e., depth), $\gamma$ is the total number of iterations, and $\mu(\cdot)$ refers to the average time per iteration. In practice, the whole algorithm is managed by a priority queue where paths are inserted and sorted according to their cost $f(n)$.

To apply the two optimizations mentioned earlier, we customize two parts of the algorithm: (i) how new nodes are inserted in the queue (sub-optimal path elimination) and (ii) how the queue is kept sorted and clean from redundant nodes (redundant nodes merging). Algorithm~\ref{alg:get_optimum} shows the pseudo-code for finding the optimal load balancing scenario with our branch-and-bound algorithm.

As discussed earlier, load balancing nodes at the same depth (i.e., iteration) are redundant. Therefore, instead of inserting load balancing nodes directly in the queue, we check if one already exists at the same iteration and replace it if the new node has a lower cumulative edge cost (line $10$ in Algorithm~\ref{alg:get_optimum} and detailed in Algorithm~\ref{alg:replaceinsert}). However, according to the sub-optimal path elimination process, we must guarantee that we can not insert a load balancing node at a given depth if the shortest path has already been discovered at this level. To do this, we implemented a lookup table in which we map the iteration (i.e., depth) to a boolean. This boolean indicates if a load balancing node has already been removed. When a new node has to be inserted, we look up inside the table to see if one has already been seen and if it does, we discard it. 

Even though the aforementioned optimizations allow to retrieve the optimal load balancing scenario, they may prune the nth best solution. Those solutions may be of interest to measure the gap between the optimal and close-to-optimal scenarios. Hence, to retrieve them as well, we need to prune fewer paths in the sub-optimal path elimination process. To recall, we previously explained that, in this process, we keep the last edge from a parent of a load balancing node only if it belongs to the shortest path from the root node to the load balancing node. This constraint has to be relaxed to allow the computation of the nth best solution. The idea is to keep the last edge from a parent of a load balancing node whether they belong \text{at least} to its nth shortest path. This has a logical meaning; in fact, if we keep all the possible edges, we end up with the original algorithm, which is able to retrieve all the solutions ordered by their cumulative edge cost. Note that the time to the solution will increase because the size of the tree increases as well.

Finally, the last point to discuss is how to find the optimal load balancing scenario in a real application when $\mathscr{C}(e)$ is measured on a real computer and not by an equation. In this setup, the idea remains the same as before. However, when a node produces its children, we computes the two edge costs by executing the corresponding iterations. Indeed, the partition and the state of the application (e.g., the position of the particles in space, their velocities, etc.) must be propagated and updated after each computation. To reduce the memory footprint, we propose to use a lookup table to store the application states as a function of their iteration. Moreover, this is necessary to guarantee that every node at the same iteration has the same application state, which is needed for results consistency. 

We made available an implementation of the optimal load balancing scenario algorithm in C++ with two different packages:
\begin{itemize}
    \item LBOPT~\cite{XetqL/LBOPT:Parameters}: This package includes the customized $A^*$ algorithm and the model presented in Section 3. LBOPT can be used to have a first idea of the performance of various load balancing criteria that can be modeled using equations. 
    \item YALBB~\cite{XetqL/yalbb:Benchmark}: It implements an N-body simulation with a short-range force. YALBB eases the benchmarking of load balancing algorithms and criteria by separating the physics from the code of interest. It employs template meta-programming and an extensive use of modern C++ constructs.
\end{itemize}

\section{Comparison of Load Balancing Criteria}
\label{sec:comparison}

In this section, we propose a comparison study of four load balancing criteria present in the literature, and we discuss their pros and cons. For that purpose, we have two approaches. First, we used synthetic benchmarks that we modeled using the equations presented in Section~\ref{sec:LBAppModel}. Therein, we compared only global load balancing criteria, and for Menon's like criteria we implemented only the original Menon criterion. For instance, we did not consider the criterion from Marquez~et~al.~\cite{MarquezClaudio2013AApplications}, because it involves the local workload from the processing elements. These synthetic benchmarks target various types of workload increase rates that create load imbalance over time. They are meant to cover as many real-life situations as possible. We studied the following schemes where the load imbalance 
\begin{itemize}
    \item Follows a linear growth, a logarithmic growth, and a quadratic growth.
    \item Auto-corrects itself periodically.
\end{itemize} 
We also performed a couple of more complex experiments with other type of load imbalance behaviors, however, these experiments require further investigations and will be part of a future work. Moreover, we used YALBB to assess the efficiency of load balancing criteria on a real-world problem. We then compared their performance against the optimum obtained using the algorithm presented in Section~\ref{sec:OptimalScenarioAlgorithm}. We employed several particle distributions and behaviors to match as closely as possible our synthetic benchmarks. 

\subsection{Synthetic benchmarks}
\label{sec:criteria-study:theory}

A parallel application is described by two main pieces of information. First, the total workload associated with the problem itself $W(t)$ (i.e., the time to compute the application on one processing unit). In case of inherently irregular applications, this workload may changes over time. Second, the distribution of the total workload among the processing elements (i.e., load imbalance) used for the computation $I(t)$. From those two pieces of information, we compute $m(t)$ and $\mu(t)$, which we use in Equation~\ref{eq:tpar_1}, to compute the application parallel time. To recall, Equation~\ref{eq:tpar_1} reads 
\begin{equation*} 
    T_{\text{par}} = \int_0^\gamma m(t) \ \text{d}t = \int_0^\gamma m(t) - \mu(t) \ \text{d}t + \int_0^\gamma \mu(t) \ \text{d}t.
\end{equation*} Using $W(t)$, we can retrieve the average workload $\mu(t)$ given a number of processing elements, whereas, we can compute $m(t)$ using the load imbalance $I(t)$ and $\mu(t)$ using the well-known percent imbalance metric~\cite{Pearce2012}
\begin{align*}
    I(t) &= \frac{m(t)}{\mu(t)} - 1,\\
    m(t) &= [I(t)-1]\mu(t).
\end{align*}  For that purpose, we have to define the function $W(t)$ and $I(t)$ and how they behave over time.

The first function, $W(t)$, gives at each iteration the total amount of work to do (expressed in units of time). It reads
\begin{equation*}
    W(t) = W_0 + \sum_{i=1}^t \omega(i),
\end{equation*} where $W_0$ is the initial application workload and $\omega(t)$ is a function giving the difference of application workload between two iterations. Hence, the average workload $\mu(t)$ is expressed as $\mu(t) = W(t) / P$. The second function, $I(t)$, gives at each iteration the load imbalance, hence it is expressed as
\begin{equation*}
    I(t) = \begin{cases}
        I(t-1) + \iota(t - \text{LB}_{\text{previous}}) \quad \text{if t } > \text{LB}_{\text{previous}}, \\
        0 \quad \text{otherwise},
    \end{cases}
\end{equation*} where LB$_{\text{previous}}$ is the previous iteration at which the load balancing mechanism has been used and $\iota(t)$ is a function returning the difference in load imbalance between two iterations. Typically, $I(t) \in [0, P-1]$, hence, in practice, a particular attention as to be given to not trespass those bounds. Now, by setting $\omega(\cdot)$, $\iota(\cdot)$, $W_0$, $P$, and the number of iterations $\gamma$, we can compute $m(t)$ and $\mu(t)$ for each iteration and use Equation~\ref{eq:tpar_1} to compute the parallel time of the application. 

\begin{remark}
Here, we make a strong assumption on the shape of the load imbalance curve after load balancing. Indeed, in practice, the load balancing mechanism involving data partitioning will influence the load imbalance growth. It is impossible to know the load imbalance function after load balancing without a deep understanding about the partitioning algorithm impact on the problem to solve, which is extremely challenging to incorporate into a mathematical model. Herein, we decide to use $\iota(t - \text{LB}_{\text{previous}})$, which means that each time a load balancing is performed, the load imbalance pattern is repeated. Another possibility could have been to set the y-intercept to $0$ (i.e., shift the function $\iota(\cdot)$ down) after each load balancing step. However, this solution was difficult to implement without providing any clear benefits. Finally, we think that this subject is worth a research effort and will be targeted in future works.
\end{remark}

\begin{table}[ht]
    \centering
    \small
\makebox[\linewidth]{
    \begin{tabular}{c|c|c|c|c|c}
    \toprule
         $\omega(t)$ & $\iota(t - \text{LB}_{\text{previous}})$ & $W_0$ & $P$ & $C$ & $\gamma$  \\
         \midrule
         $0$ & $0.1$ & $52*P$ & $10{,}649{,}600$ & $W_0*P*10^2$ & $600$ \\
         $0$ & $1/(0.4 * t + 1)$ & $52*P$ & $10{,}649{,}600$ & $W_0*P*10^2$ & $600$  \\
         $0$ & $0.02 * t$ & $52*P$ & $10{,}649{,}600$ & $W_0*P*10^2$ & $600$  \\
         $0$ & $-(0.1 * (t\% 17)) + 0.8$  & $52*P$ & $10{,}649{,}600$ & $W_0*P*10^2$ & $600$  \\
         \midrule
         $\sin{\frac{\pi t}{180}}$ & $0.1$ & $52*P$ & $10{,}649{,}600$ & $W_0*P*10^2$ & $600$ \\
         $\sin{\frac{\pi t}{180}}$  & $1/(0.4 * t + 1)$ & $52*P$ & $10{,}649{,}600$ & $W_0*P*10^2$ & $600$  \\
         $\sin{\frac{\pi t}{180}}$  & $0.02 * t$ & $52*P$ & $10{,}649{,}600$ & $W_0*P*10^2$ & $600$  \\
         $\sin{\frac{\pi t}{180}}$  & $-(0.1 * (t\% 17)) + 0.8$  & $52*P$ & $10{,}649{,}600$ & $W_0*P*10^2$ & $600$  \\
    \bottomrule
    \end{tabular}
    }
    \caption{Parameters used to define the synthetic benchmarks. Two types of situations have been considered. The first one (top side of the table) considers benchmarks with a static workload and irregular workload distribution, whereas the second one (bottom side of the table) targets a benchmark with an irregular workload and an irregular workload distribution. All workload are expressed in time units.}
    \label{tab:synthetic_benchmarks}
\end{table} The parameters used in the synthetic benchmarks are summarized in Table~\ref{tab:synthetic_benchmarks}. The initial workload (expressed in time units) is proportional to a $2$D Lattice-Boltzmann computational fluid dynamic problem with $10^9$~D2Q9~cells per processing unit  with a performance of $1$~Gflops~\cite{Tomczak2018SparseProcessors}. The number of processing units is equal to the number of cores available in the supercomputer ``Sunway TaihuLight''~\cite{Top5002020}. We studied two types of situations. First, we targeted benchmarks with a static workload (i.e., the global workload is always the same) but with a workload distribution that changes over time. Then, we focused on the same benchmarks but with an irregular workload that increases/decreases over time. The static workload benchmarks target applications that suffer from load imbalance due to the parallelization. The irregular workload benchmarks target applications with varying workload per time-step, and where the load imbalance comes from both the problem in itself and the parallelization. 


\begin{figure}[H]
\begin{center}
\subfloat[The load imbalance increase rate is constant: $\iota(t) = 0.1$\label{fig:benchmark:static:constant}] {

\includegraphics[width=.45\linewidth]{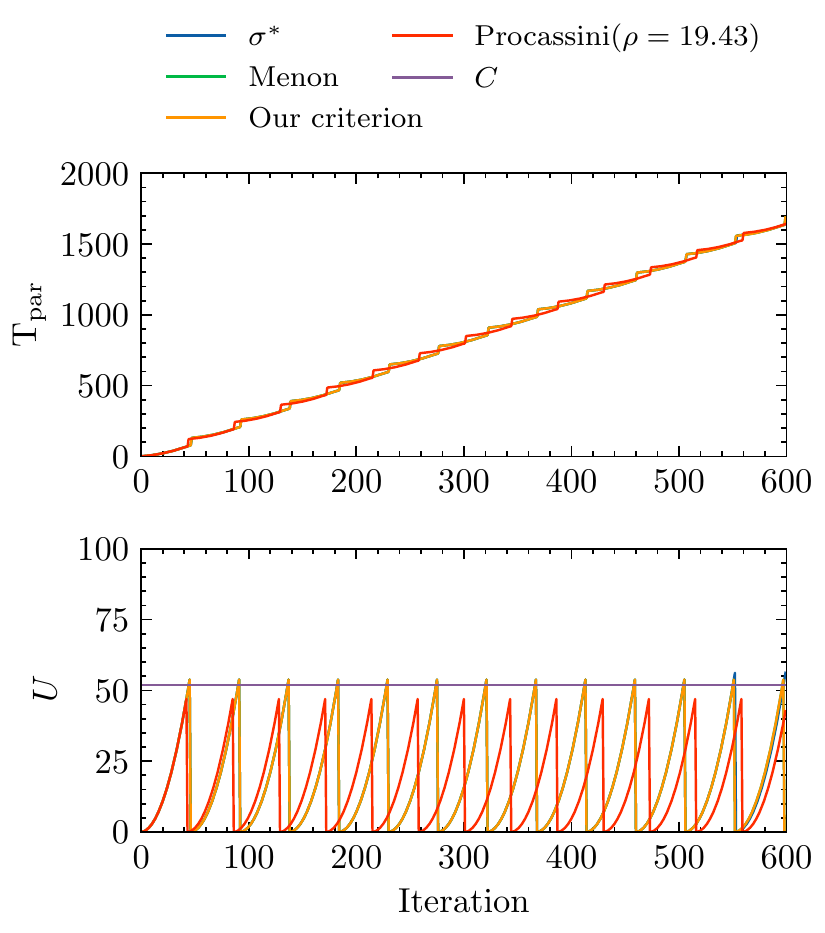}
}\quad
\subfloat[The load imbalance increase rate is linear with time: $\iota(t) = 0.02t$\label{fig:benchmark:static:linear}]      {
\includegraphics[width=.45\linewidth]{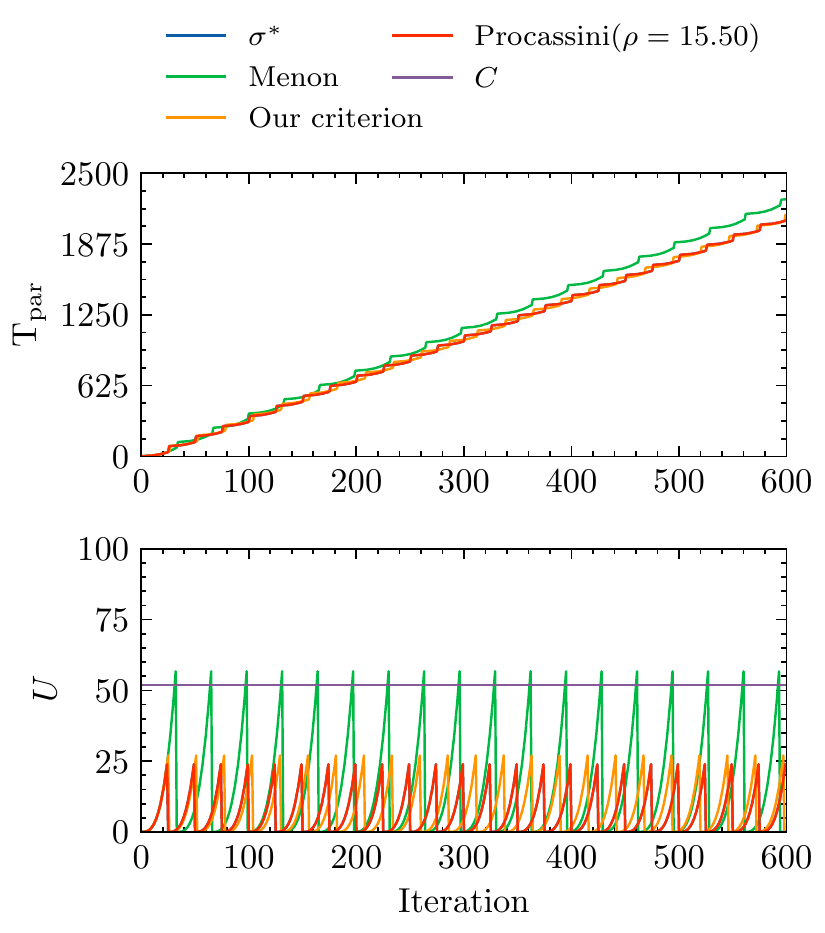}
}\\
\subfloat[The load imbalance increase rate is sub-linear with time: $\iota(t) = \frac{1}{0.4t+1}t$\label{fig:benchmark:static:sublinear}]      {
\includegraphics[width=.45\linewidth]{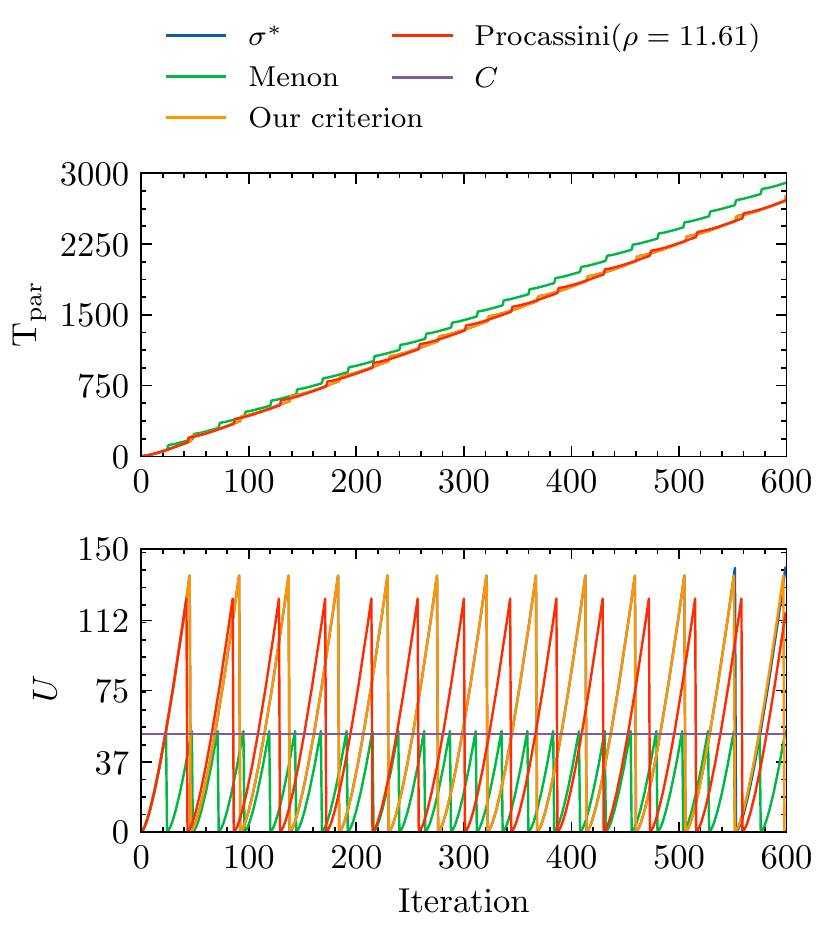}
}\quad
\subfloat[Load imbalance corrects itself every $17$ iterations: $\iota(t) = -0.1*(t\%17) + 0.8$\label{fig:benchmark:static:autocorrect}]      {
\includegraphics[width=.45\linewidth]{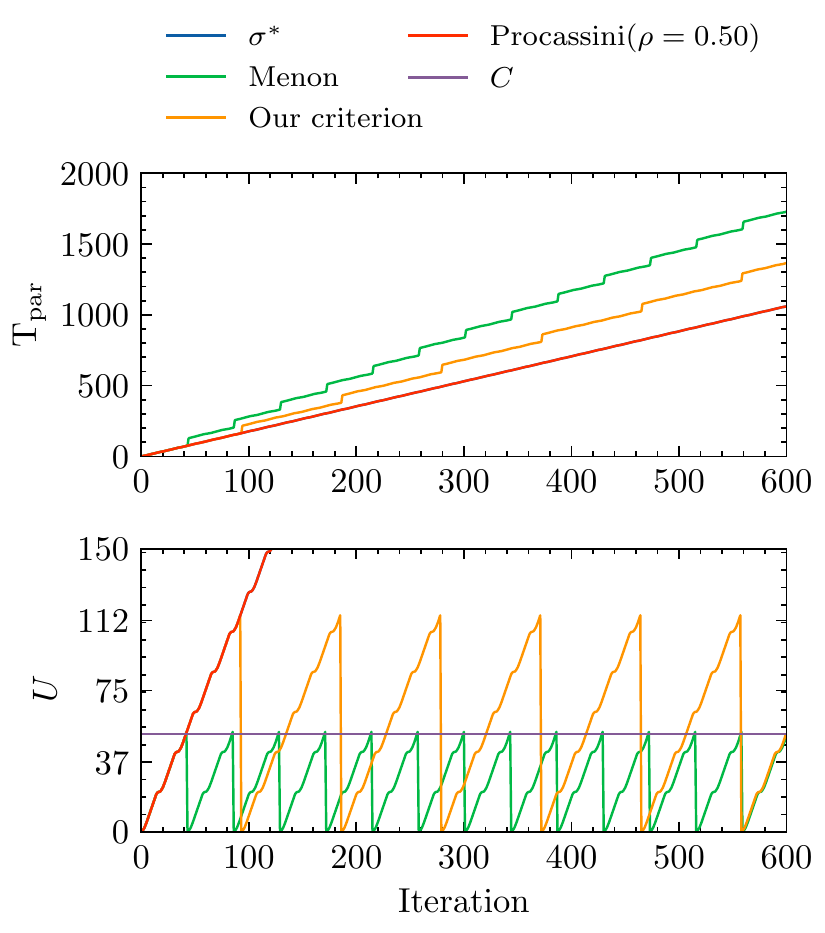}
}
\end{center}
\caption{Results of the synthetic benchmarks with static workloads for Menon criterion~\cite{H.MenonandN.JainandG.ZhengandL.Kale2012}, our criterion, and Procassini criterion~\cite{Procassini2004LoadCalculations} against the optimal scenario $\sigma^*$. Four sources of load imbalance are considered: (i) constant, (ii) linear with time, (iii) sublinear with time, and (iv) linear with time and self-correcting every $17$ iterations. Parameters used in this benchmark are summarized in Table~\ref{tab:synthetic_benchmarks}. $U$ is defined in Equation~\ref{eq:U}.}
\label{fig:benchmark:static}
\end{figure}

The results of the synthetic benchmarks for static applications are shown in Figure~\ref{fig:benchmark:static}. For Procassini criterion, we tried $5000$ values of $\rho$ between $0.5$ and $50.0$; however, for readability, we decided only to show the scenario that performed the best. The upper figure shows the simulated parallel time that we obtained using the model presented in Section~\ref{sec:LBAppModel}. The lower figure indicates the growth of the cumulative time-per-time-step $U$ (defined in Equation~\ref{eq:U}), and the horizontal bar gives the value of $C$ in order to track how the criteria differ from Menon criterion given by Equation~\ref{eq:MenonCrit}. We use this figure to compare the behavior of the load balancing criteria.

In the constant experiment (Figure~\ref{fig:benchmark:static:constant}), both Menon criterion and our criterion behave like the optimal strategy. In other words, their load balancing time interval is similar to $\sigma^*$ . Still, they differ marginally at the end of the simulation. It is important to remark that depending on when the last load balancing step happens, it may be preferable to delay or schedule some load balancing calls earlier, as we can observe in Figure~\ref{fig:benchmark:static:constant}. Indeed, wasting a call at the very end of a simulation is useless. However, to take such a decision, one may need to foresee the future and detect if, given the current criterion, a call would appear near the end. Obviously, only the solution from our branch-and-bound algorithm is able to see that, as it tests ``all'' the possible solutions. Procassini criterion with a $\rho$ value of $19.43$ seems optimal. Note that we also tried to use $\rho_\tau$ (defined in Equation~\ref{eq:rhotau}) for Procassini criterion. We observed that Procassini criterion performs the load balancing steps at the exact same iteration than Menon and our criterion, as suggested in Remark~\ref{rem:proca=menon=ours}. Finally, this experiment fits well the hypothesis of both our criterion and Menon criterion, and thus they behave optimally, as shown in Section~\ref{sec:LBAppModel}. 

In the linear experiment (Figure~\ref{fig:benchmark:static:linear}), our criterion and Procassini criterion with a $\rho$ value of $15.5$ behave like the optimal scenario and therefore are very close in terms of performance. However, we notice that Menon criterion does not follow the same load balancing time interval as the optimal scenario, leading to a performance loss of approximately $10\%$. In particular, we remark that Menon criterion does not re-balance frequently enough.

In the sublinear experiment (Figure~\ref{fig:benchmark:static:sublinear}), the opposite situation appears (compared to the linear experiment). Herein, Menon criterion re-balance too often, wasting useful resources. This is an expected behavior as we observed in Section~\ref{sec:LBAppModel} that this criterion is optimal only if the load imbalance growth is constant, which is not the case in the sublinear experiment nor in the linear experiment.  

In the auto-correct experiment (Figure~\ref{fig:benchmark:static:autocorrect}), we see that neither our criterion nor Menon criterion can understand that no load balancing is required because the load imbalance corrects itself periodically. Nevertheless, our criterion is able to detect up to five auto-correcting patterns in a row. For that reason, in this experiment, our criterion is far better in terms of performance than Menon criterion. Only Procassini criterion provided the optimal $\rho$ value is able to match the performance of the optimal scenario.

\begin{figure}[H]
\begin{center}
\subfloat[The load imbalance increase rate is constant: $\iota(t) = 0.1$\label{fig:benchmark:irregular:constant}]      {

\includegraphics[width=.45\linewidth]{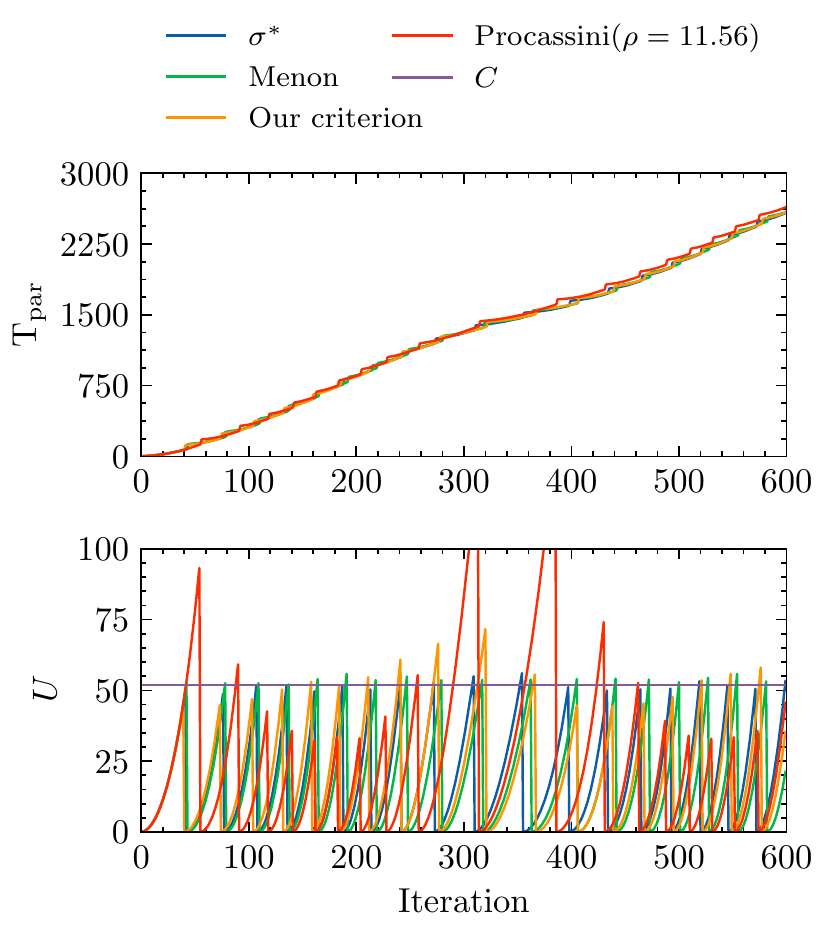}
}\quad
\subfloat[The load imbalance increase rate is linear with time: $\iota(t) = 0.02t$\label{fig:benchmark:irregular:linear}]      {
\includegraphics[width=.45\linewidth]{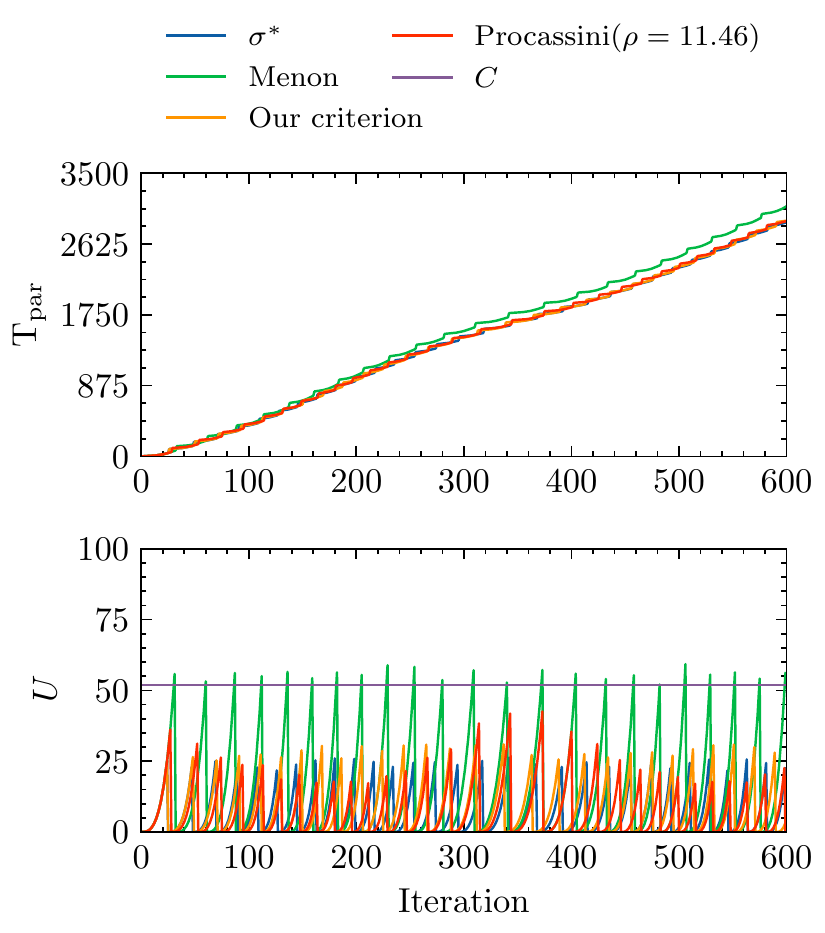}
}\\
\subfloat[The load imbalance increase rate is sub-linear with time: $\iota(t) = \frac{1}{0.4t+1}t$\label{fig:benchmark:irregular:sublinear}]      {
\includegraphics[width=.45\linewidth]{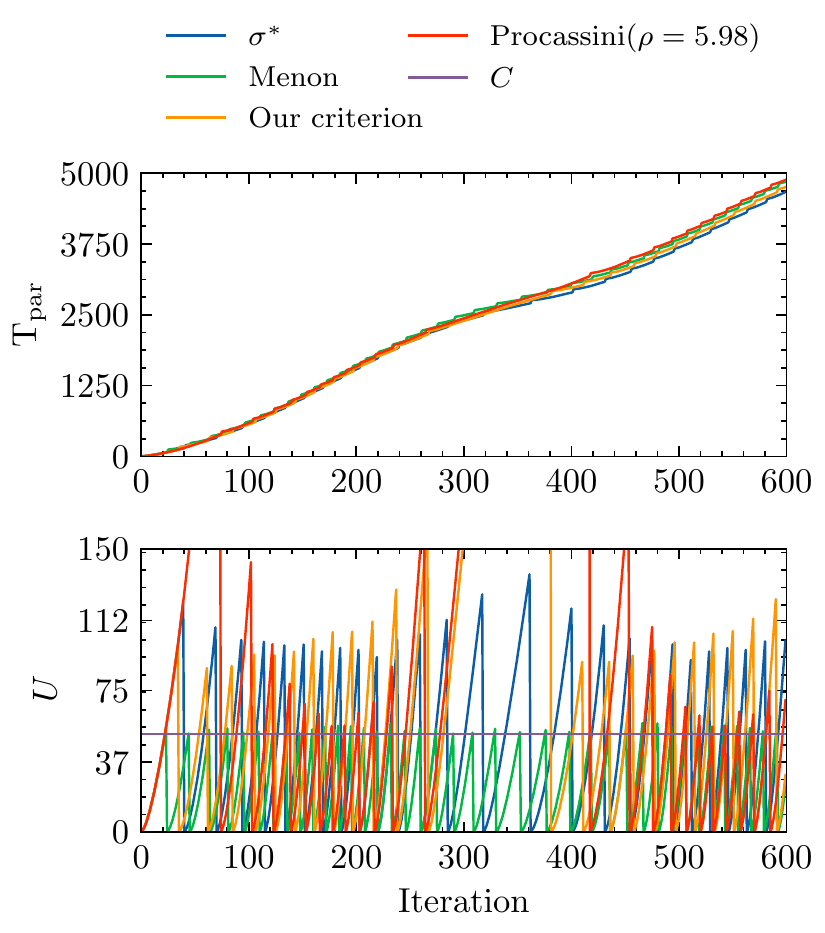}
}\quad
\subfloat[Load imbalance corrects itself every $17$ iterations: $\iota(t) = -0.1*(t\%17) + 0.8$\label{fig:benchmark:irregular:autocorrect}]      {
\includegraphics[width=.45\linewidth]{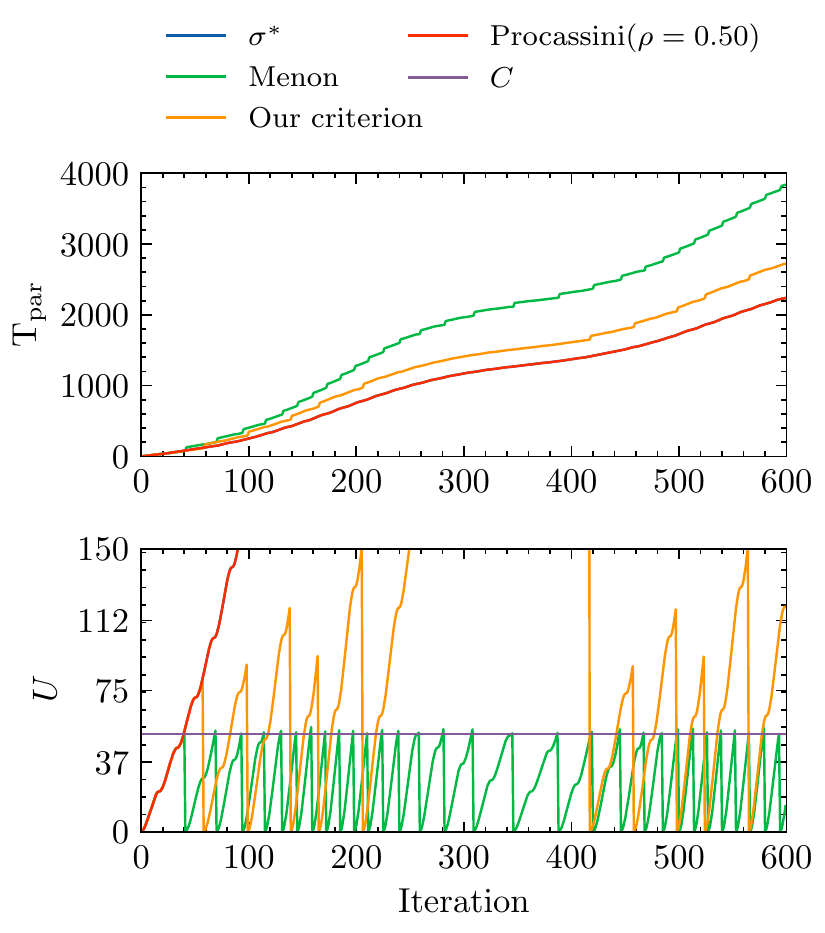}
}
\end{center}
\caption{Results of the synthetic benchmarks with irregular workloads for Menon criterion~\cite{H.MenonandN.JainandG.ZhengandL.Kale2012}, our criterion, and Procassini criterion~\cite{Procassini2004LoadCalculations} with an optimally tuned parameter $\rho$ against the optimal scenario $\sigma^*$. Four sources of load imbalance are considered: (i) constant, (ii) linear with time, (iii) sublinear with time, and (iv) linear with time and self-correcting every $17$ iterations. Parameters used in this benchmark are summarized in Table~\ref{tab:synthetic_benchmarks}. $U$ is defined in Equation~\ref{eq:U}.} 
\label{fig:benchmark:irregular}
\end{figure}

The results of the synthetic benchmarks with irregular workloads are presented in Figure~\ref{fig:benchmark:irregular}. Like in the previous benchmarks, the same values of $\rho$ have been considered for Procassini criterion, and we show the scenario that performed the best. 

In the constant experiment presented in Figure~\ref{fig:benchmark:irregular:constant}, we see that the performance of both our criterion and Menon criterion are almost unchanged. However, Procassini criterion decreased in performance compared to the static experiment.  In the linear experiment (Figure~\ref{fig:benchmark:irregular:linear}), the results are similar to the static experiment where Menon criterion does not re-balance frequently enough, whereas our criterion and Procassini criterion follow the behavior of the optimal strategy. 

In the sub-linear experiment (Figure~\ref{fig:benchmark:irregular:sublinear}), Menon criterion improves its performance, whereas Procassini criterion's performance decreases. Our criterion performs on par with both Menon criterion and the optimal scenario. It is worth noticing that during the slow-down around iteration $300$, our criterion stops re-balancing while the optimal scenario only decreases the load balancing time interval. This suggest that our criterion is able to adapt its behavior to the current situation. This phenomenon is also visible in the last experiment. However, there, the optimal scenario does not re-balance at all. Finally, in the auto-correct experiment (Figure~\ref{fig:benchmark:irregular:autocorrect}), we remark that Procassini criterion is the only criterion able to detect that re-balancing the application is not necessary. Nevertheless, our criterion reduces its load balancing time interval, drastically improving its performance compared to Menon criterion.

\begin{figure}[ht!]
\begin{center}
\subfloat[Static workload\label{fig:relperf_static}]      {
\includegraphics[width=.45\linewidth]{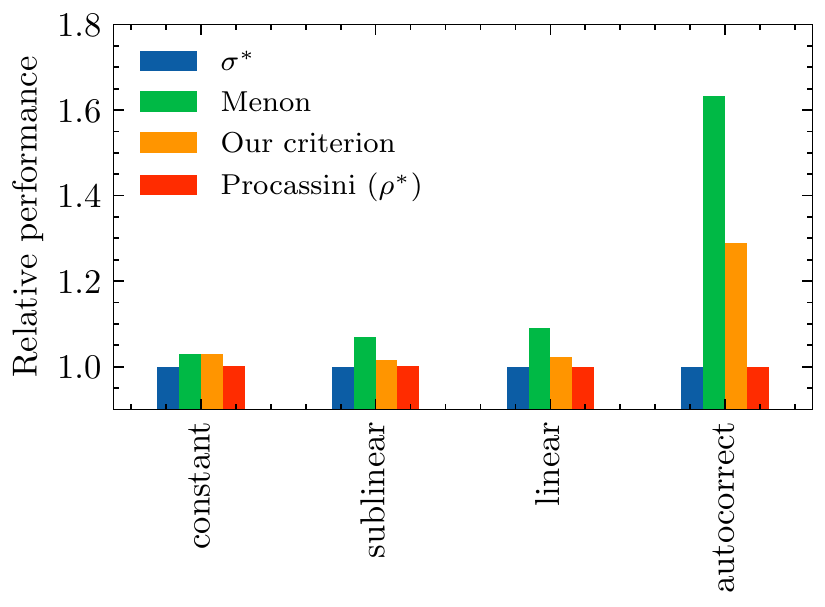}
}\quad
\subfloat[Irregular workload\label{fig:relperf_irregular}]      {
\includegraphics[width=.45\linewidth]{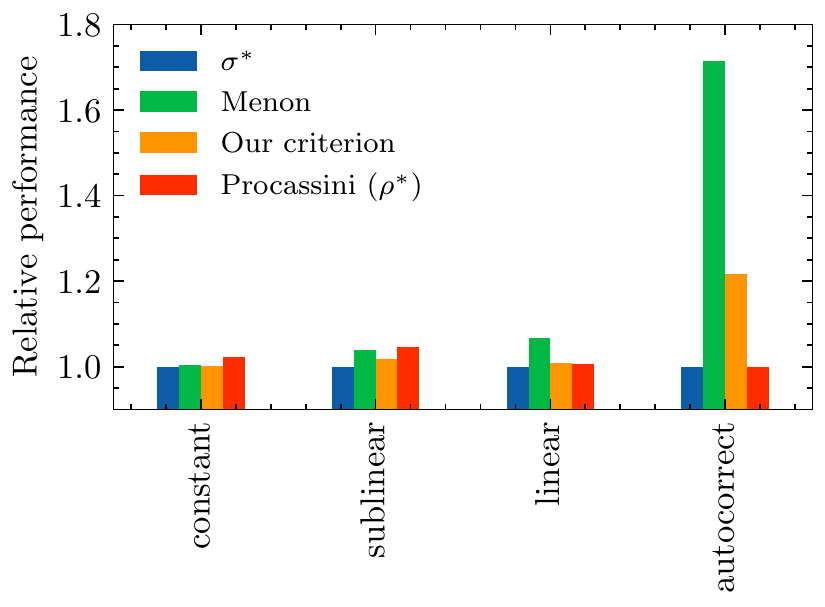}
}
\end{center}
\caption{Relative performance of the our criterion, Menon criterion, and Procassini criterion against the optimal scenario in the static workload and irregular workload synthetic benchmarks. The relative performance is defined as $T_{\text{criteria}} / T_{\sigma^*}$.}
\label{fig:relative_performance_static}
\end{figure}

To understand the difference in performance among those criteria in a better way, we show in Figure~\ref{fig:relative_performance_static} the relative performance of our criterion, Menon criterion, and Procassini criterion compared to the optimal scenario. The relative performance is defined as $T_{\text{criteria}} / T_{\sigma^*}$. We see that out of the three criteria we studied Procassini criterion is the best provided the optimal value of $\rho$. However, not every scientist can afford effort to find the optimal $\rho$ before executing his/her application, which is not needed with our criterion and Menon's like criteria. Moreover, the performance of both Menon criterion and our criterion are really close to the optimal scenario in these experiments. Finally, Menon criterion performs better in the irregular workload than in the static workload situations.

To confirm these hypotheses, we propose a numerical study of all the criteria presented in Section~\ref{section:background_work} on YALBB, a home-made load balancing benchmark based on a N-body simulation with a short-range force. 

\subsection{Numerical study with YALBB}
\label{sec:criteria-study:numerical}

\begin{figure}[ht!]
    \centering
    \includegraphics[width=0.5\textwidth]{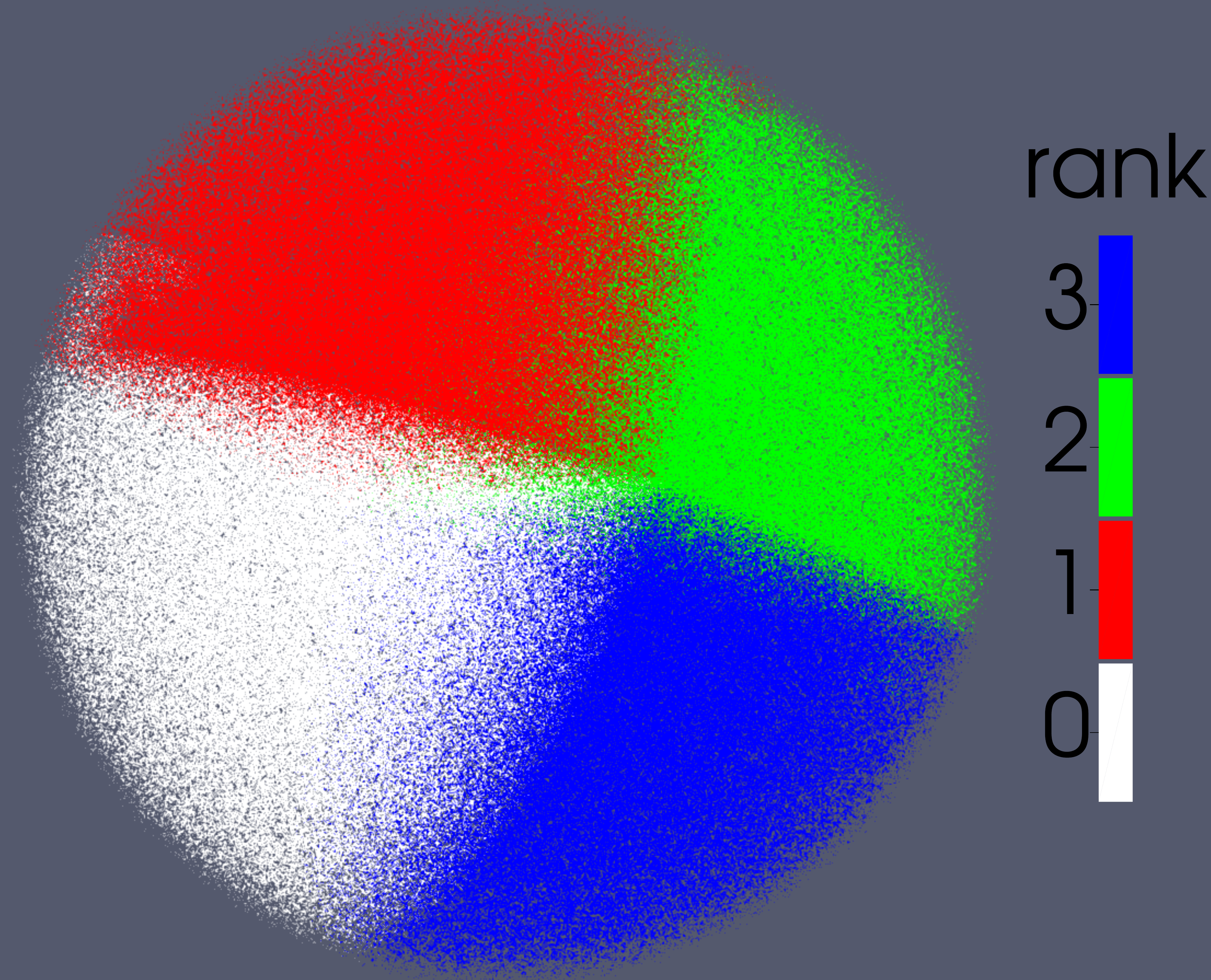}
    \caption{Example of a sphere of uniformly distributed $40{,}000$ particles. Particles are distributed of $4$ processing elements using an Hilbert~Space~Filing~Curve. The particles are colored according to the rank of their attributed processing element.}
    \label{fig:yalbb:example} 
\end{figure}

 We carried out three experiments involving $40{,}000$ particles and hundreds of millions of interactions with ``YALBB'' to evaluate the load balancing criteria presented in Section~\ref{section:background_work}.
The experiments were conducted with a standard Lennard-Jones interaction. The inner data structure uses the well-known cell lists algorithm for managing particles neighborhood. In these experiments, we used Zoltan~\cite{Devine2002} as a load balancing library for partitioning and managing the related data. Figure~\ref{fig:yalbb:example} shows an example of $40{,}000$ particles distributed among $4$ processing elements using the Hilbert space-filling curve algorithm available in Zoltan. The experiments were executed on ``Yggdrasil'' the University of Geneva's cluster (Intel Xeon Gold 6240 CPU @ 2.60GHz).

\begin{figure}[ht!]
\begin{center}
\subfloat[The sphere of particles is contracting, the number of interactions grows.\label{fig:numerical:workload:contraction}]      {
\includegraphics[width=.45\linewidth]{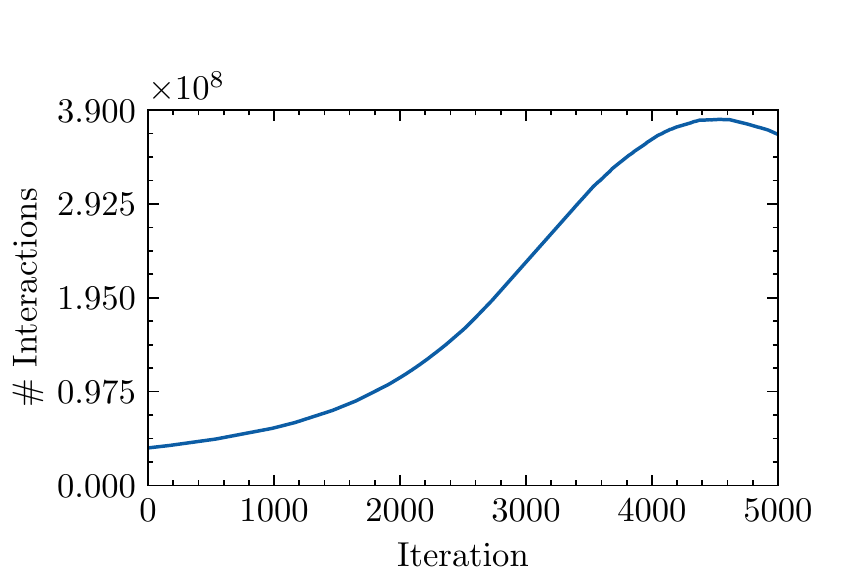}
}\quad
\subfloat[The sphere of particles is expanding, the number of interactions decreases.\label{fig:numerical:workload:expansion}]      {
\includegraphics[width=.45\linewidth]{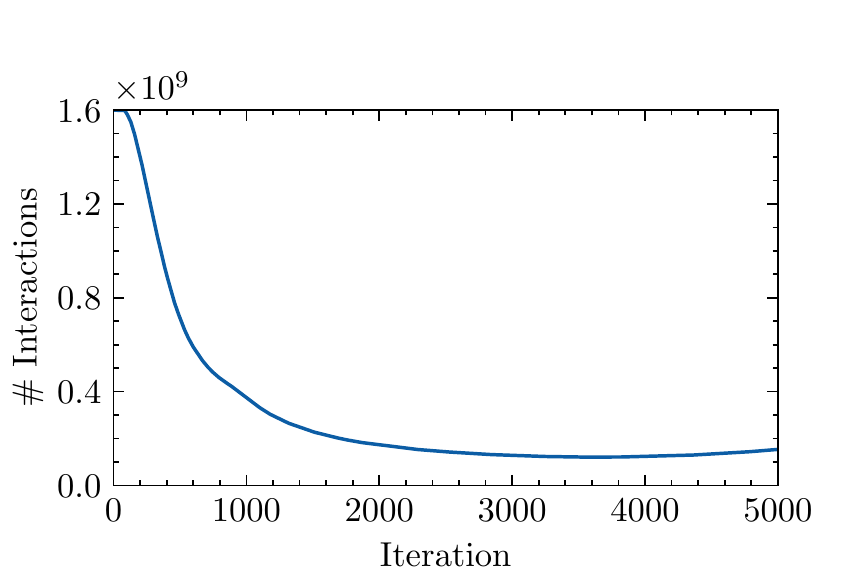}
}\quad
\subfloat[The sphere of particles is expanding and then contracting, the number of interactions decreases and then increases.\label{fig:numerical:workload:expansion-contraction}] {
\includegraphics[width=.45\linewidth]{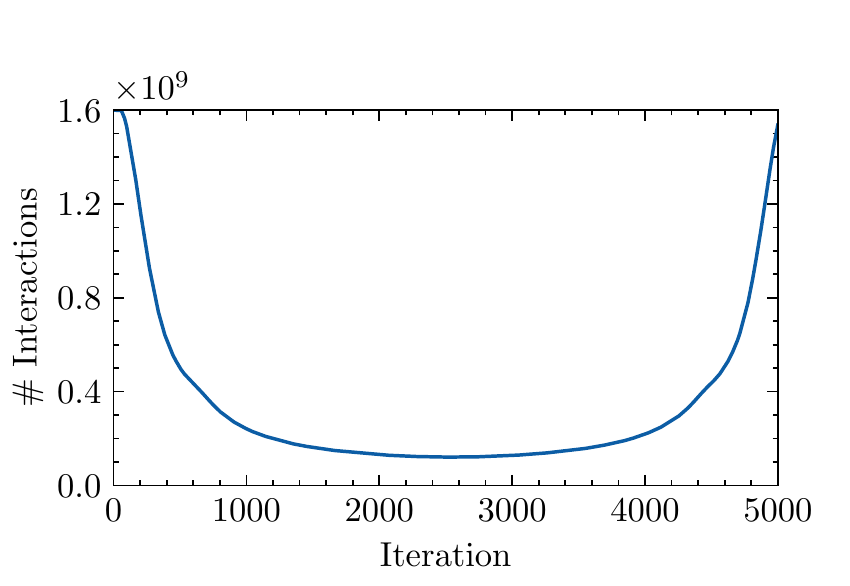}
}\end{center}
\caption{The number of particle interactions to compute at each iteration (i.e., application workload) of the three experiments carried out in the numerical study. Each experiment is composed of $40{,}000$ particles. The first experiment computes a sphere of uniformly distributed particles that contracts on the effect of an external force. The second experiment computes a sphere of uniformly distributed particles that expands. The third experiment starts by expanding the sphere and then the sphere contracts.}
\label{fig:numerical:interactions}
\end{figure}

The first experiment consisted of a uniformly distributed sphere of particles expanding in a vacuum. The second experiment simulated the compression of a bigger uniformly distributed sphere of particles in a vacuum. The third experiment was a combination of both, one after the other, starting with the expansion phase. While in expansion, the particles were attracted to the center of the sphere by a force proportional to the earth's gravity. Hence, after a few iterations, the sphere started to compress again. The results are obtained over one sequence of expansion-compression of the gas. 
\begin{table}[h]
\begin{adjustbox}{center}
\begin{tabular}{lccc}
\toprule
 Parameter                & Contraction & Expansion & Expansion and Contraction \\
\midrule
Box size (x,y,z) & \multicolumn{3}{c}{($3.15$, $3.15$, $3.15$)}\\
Number of particles & \multicolumn{3}{c}{$40{,}000$}\\
$\sigma_{\text{LJ}}$ & \multicolumn{3}{c}{$0.7$}\\
$\epsilon_{\text{LJ}}$ & \multicolumn{3}{c}{$1.0$}\\
Initial temperature & \multicolumn{3}{c}{$3.0$}\\
Time-step & $2e$-$5$ & $8.4e$-$5$ & $1.2e$-$4$\\
\bottomrule
\end{tabular}
\end{adjustbox}
\caption{Physical parameters for the three numerical experiments.}
\label{tab:numerical:phyparams}
\end{table}
The physical parameters used in our numerical study are shown in Table~\ref{tab:numerical:phyparams}. The number of interactions to compute over time is shown in Figure~\ref{fig:numerical:interactions} for each experiment. As we can see in this figure, the amount of interactions (i.e., the density of particles) varies a lot over the execution of the experiment changing the requirement for load balancing. At the beginning of the expansion simulations, almost every particle interacts with all the others, this huge density decreases rapidly after the beginning of the simulation, drastically changing the workload of many processing elements. The reverse situation appears in the contraction simulation where there is almost no interaction at the beginning of the code execution, but a very high density is observed towards the end. In these experiments, we used the Hilbert Space Filing curve algorithm available in the Zoltan load balancing library~\cite{Devine2002}.

\begin{figure}[ht!]
    \centering
    \includegraphics[width=0.85\textwidth]{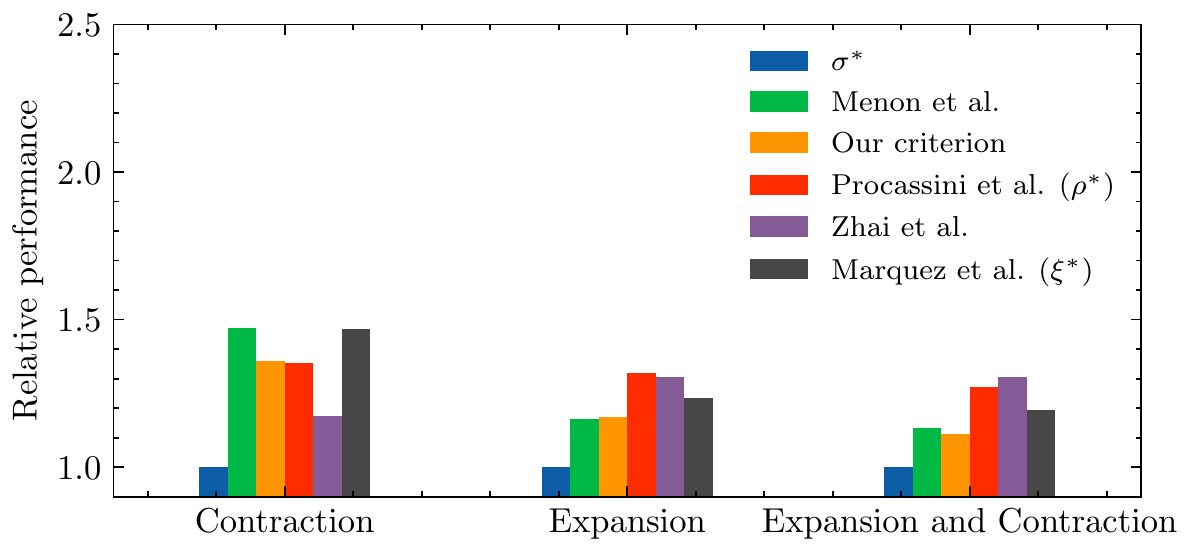}
    \caption{Comparison of the best performance of each criterion (among $5$ executions) in the numerical experiments relative to the optimal scenario $\sigma^*$. }
    \label{fig:yalbb:performance}
\end{figure}

The results of the three experiments are presented in Figure~\ref{fig:yalbb:performance}. We executed the code $5$ times for each experiment, and we took the best parallel time for each criterion. As we can observe, state-of-the-art load balancing criteria can achieve close to optimal performance. However, for Procassini criterion and Marquez criterion, the user has to find the optimal value of the parameter ($\rho$ or $\xi$), which is not something everybody can afford. This is why automatic criteria seem to be the best fit for most situations, even though a $\rho$ value between $1.0$ and $1.25$ seems to work the best for Procassini criterion. Furthermore, as we can see in Table~\ref{tab:numerical:result_param}, criteria with an extra parameter often have non-consistent results across experiments. Also, the penalty for using a sub-optimal value can be huge, and there is no rule of thumb to find the right value except testing many of them. Finally, we see that our criterion performs, on average, $4.9\%$ faster than the other load balancing criteria. In particular, our criterion is $17.64\%$ faster than Zhai criterion in the expansion-contraction experiment. 

\begin{table}[ht!]
\begin{adjustbox}{center}
\begin{tabular}{llll}
 \toprule
 Experiment                & Criterion         & Best [s]                     & Worst [s]                     \\
 \midrule
 \multirow{6}{*}{Contraction} & $\sigma^*$        & 20.6500  & -   \\
& Menon et al.      & 30.4383  & -            \\
& Our criterion               & 28.0711  & -          \\
& Zhai et al.       & 24.6481  & -           \\
& Procassini et al. & 28.4504 ($\rho = 1.25$) & 394.227 ($\rho = 15.00$) \\
& Marquez et al.    & 30.3177 ($\xi = 4.00$)  & 196.65 ($\xi = 0.50$)       \\
 \midrule
  \multirow{6}{*}{Expansion} & $\sigma^*$        & 20.3852  & -    \\
& Menon et al.      & 23.7408  & -           \\
& Our criterion               & 23.8811  & -          \\
& Zhai et al.       & 26.6061  & -           \\
& Procassini et al. & 26.9306 ($\rho = 1.00$) & 253.109 ($\rho = 10.00$) \\
& Marquez et al.    & 25.2197 ($\xi = 0.90$)  & 37.0333 ($\xi = 0.50$)      \\
 \midrule
 \multirow{6}{*}{Expansion and Contraction}  & $\sigma^*$        & 25.6118  & -   \\
& Menon et al.      & 28.9729  & -           \\
& Our criterion     & 28.4663  & -           \\
& Zhai et al.       & 33.4903  & -           \\
& Procassini et al. & 32.5988 ($\rho = 1.00$) & 270.571 ($\rho = 10.00$) \\
& Marquez et al.    & 30.5376 ($\xi = 1.50$)  & 54.5368 ($\xi = 0.50$)      \\
 \bottomrule
\hline
\end{tabular}
\end{adjustbox}
\caption{Summary of the performance results for the three numerical experiments. The performance of each criterion (and for each extra parameter value) is based on $5$ executions. For the criteria with an extra parameter, the parameter value that lead to the best and worst performance is shown between parenthesis.}
\label{tab:numerical:result_param}
\end{table}
Among Menon's like criteria, the Zhai criterion seems to be the less stable one. Even though it outperforms Menon criterion in one experiment, the Zhai criterion produced a run considerably slower in the other two experiments. This suggests that different implementation of the same idea behind load balancing criteria might have a huge impact on performance. 

Finally, we observe that our criterion performs on par with Menon criterion and outperforms it in the expansion and contraction simulation. As seen in the synthetic benchmarks, Menon criterion seems to perform better when the application exhibits an irregular workload. This could be why the gap between the two criteria is much closer in the numerical experiment than in the synthetic benchmark. Overall, our criterion and Menon criterion seem to be the most stable criteria. They are respectively only $35.93\%$ and $47.36\%$ slower than the optimum in the contraction experiment, $17.14\%$ and $16.46\%$ slower than the optimum in the expansion experiment, and $11.14\%$ and $13.12\%$ slower than the optimum in the expansion-contraction experiment. 

The present study is not enough to conclude that our criterion is better than Menon criterion, even though our criterion outperforms Menon criterion up to $8.4\%$ while it was slower by only at most $0.5\%$. However, this suggests that they are both good alternatives. Especially, this numerical study indicates that these two criteria constantly perform almost optimally. Therefore, we encourage scientists to use our branch-and-bound algorithm to compare the performance of available load balancing criteria to assess which criterion is the most suited for their type of problem. 

\section{Conclusion}
\label{sec:conclusion}

In the present paper, we proposed a review of state-of-the-art load balancing criteria and we introduced a novel fully automatic criterion based on a simple mathematical model inspired from the literature. We tried to classify these criteria as a function of their requirements and the information (external or not) required to compute the load balancing decision. Secondly, we proposed a branch-and-bound method for computing the set of load balancing steps leading to the optimal performance of a given application. Besides, we provide two implementations of this algorithm. The first implementation in LBOPT~\cite{XetqL/yalbb:Benchmark} the package related to the synthetic benchmarks and the second, in YALLB~\cite{XetqL/LBOPT:Parameters} the package related to the N-body solver we used in the numerical experiments. Afterward, we studied the performance of state-of-the-art load balancing criteria as well as our new criterion on synthetic benchmarks (modeled via our simple mathematical model) and on a parallel N-body solver. 

We observed that our novel criterion outperforms automatic state-of-the-art criteria in synthetic benchmarks. However, we pointed out that the performance difference was tighter in the irregular total workload scheme compared to the static total workload scheme. We also identified that modeling the impact of the load balancing method on the load imbalance growth is challenging. This is a topic that is worth the research effort and will be targeted for future work. 

We saw that the gain of our criterion with respect to the other criteria is much smaller in our N-body numerical experiments. Moreover, we remarked that fully automatic criteria have more reliable results, only at most $47.36\%$ (Menon criterion) slower than the optimal scenario. In particular, a run with our criterion is never more than $35.93\%$ slower than the optimum. Moreover, our criterion can outperform Menon criterion by up to $8.4\%$, while it is outperformed by up to a marginal $0.5\%$ in the worst case. We also noticed that our criterion is, on average, $4.9\%$ faster than the other load balancing criteria. All these experiments suggest that our criterion is a very good alternative to automatic load balancing criteria, offering almost optimal performance.  

Of course, to further confirm the aforementioned observations, we plan to test our new criterion on production codes. The first step will be to integrate our re-balancing strategy in Palabos~\cite{Latt2021Palabos:Solver}, a parallel Lattice-Boltzmann solver. Then, we will investigate more complex load imbalance growth. For instance, we plan to add random bias to the load imbalance growth to simulate perturbations coming from various sources. The last step is to improve our understanding about the impact of the partitioning method on the load imbalance growth. This would be useful to have benchmarks that better reproduce the behavior of real applications.  

\section{Acknoledgments}
We want to thank Olivier Belli for his help in proofreading the paper. 

\bibliographystyle{./elsarticle-num.bst}
\bibliography{main}

\begin{thebibliography}{10}
\expandafter\ifx\csname url\endcsname\relax
  \def\url#1{\texttt{#1}}\fi
\expandafter\ifx\csname urlprefix\endcsname\relax\def\urlprefix{URL }\fi
\expandafter\ifx\csname href\endcsname\relax
  \def\href#1#2{#2} \def\path#1{#1}\fi

\bibitem{Pearce2014}
O.~T. Pearce, M.~L. Adams, B.~R. De~Supinski, L.~Rauchwerger, V.~E. Taylor,
  {Load Balancing Scientific Applications}, Ph.D. thesis, Texas A{\&}M
  University (2014).

\bibitem{Garey1979}
M.~R. Garey, D.~S. Johnson, {Computers and Intractability: A Guide to the
  Theory of NP-Completeness}, W.H. Freeman {\&} Co., New York, USA, 1979.

\bibitem{Simon1997HowBisection}
H.~D. Simon, S.~H. Teng, {How good is recursive bisection?}, SIAM Journal of
  Scientific Computing 18~(5) (1997) 1436--1445.
\newblock \href {https://doi.org/10.1137/S1064827593255135}
  {\path{doi:10.1137/S1064827593255135}}.

\bibitem{Borrell2019ParallelBalancing}
R.~Borrell, G.~Oyarzun, D.~Dosimont, G.~Houzeaux, {Parallel SFC-based mesh
  partitioning and load balancing}, in: Proceedings of ScalA 2019: 10th
  Workshop on Latest Advances in Scalable Algorithms for Large-Scale Systems -
  Held in conjunction with SC 2019: The International Conference for High
  Performance Computing, Networking, Storage and Analysis, Institute of
  Electrical and Electronics Engineers Inc., 2019, pp. 72--78.
\newblock \href {https://doi.org/10.1109/ScalA49573.2019.00014}
  {\path{doi:10.1109/ScalA49573.2019.00014}}.

\bibitem{VanDriessche1995AnBalancing}
R.~Van~Driessche, D.~Roose, {An improved spectral bisection algorithm and its
  application to dynamic load balancing}, Parallel Computing 21~(1) (1995)
  29--48.
\newblock \href {https://doi.org/10.1016/0167-8191(94)00059-J}
  {\path{doi:10.1016/0167-8191(94)00059-J}}.

\bibitem{Karypis1998AGraphs}
G.~Karypis, V.~Kumar, {A fast and high quality multilevel scheme for
  partitioning irregular graphs}, SIAM Journal of Scientific Computing 20~(1)
  (1998) 359--392.
\newblock \href {https://doi.org/10.1137/S1064827595287997}
  {\path{doi:10.1137/S1064827595287997}}.

\bibitem{Fleissner2008}
F.~Fleissner, P.~Eberhard, {Parallel load-balanced simulation for short-range
  interaction particle methods with hierarchical particle grouping based on
  orthogonal recursive bisection}, International Journal for Numerical Methods
  in Engineering 74~(4) (2008) 531--553.
\newblock \href {https://doi.org/10.1002/nme.2184}
  {\path{doi:10.1002/nme.2184}}.

\bibitem{Fattebert2012}
J.-L. Fattebert, D.~F. Richards, J.~N. Glosli, {Dynamic load balancing
  algorithm for molecular dynamics based on Voronoi cells domain
  decompositions}, Computer Physics Communications 183 (2012) 2608--2615.
\newblock \href {https://doi.org/10.1016/j.cpc.2012.07.013}
  {\path{doi:10.1016/j.cpc.2012.07.013}}.

\bibitem{Furuichi2017IterativeInteractions}
M.~Furuichi, D.~Nishiura, {Iterative load-balancing method with multigrid level
  relaxation for particle simulation with short-range interactions}, Computer
  Physics Communications 219 (2017) 135--148.
\newblock \href {https://doi.org/10.1016/j.cpc.2017.05.015}
  {\path{doi:10.1016/j.cpc.2017.05.015}}.

\bibitem{Boulmier2019OnApplications}
A.~Boulmier, F.~Raynaud, N.~Abdennadher, B.~Chopard, {On the Benefits of
  Anticipating Load Imbalance for Performance Optimization of Parallel
  Applications}, in: Proceedings - IEEE International Conference on Cluster
  Computing, ICCC, Vol. 2019-September, Institute of Electrical and Electronics
  Engineers Inc., 2019.
\newblock \href {https://doi.org/10.1109/CLUSTER.2019.8890998}
  {\path{doi:10.1109/CLUSTER.2019.8890998}}.

\bibitem{Pearce2012}
O.~Pearce, T.~Gamblin, B.~R. de~Supinski, M.~Schulz, N.~M. Amato, {Quantifying
  the effectiveness of load balance algorithms}, in: Proceedings of the 26th
  ACM international conference on Supercomputing - ICS '12, ACM Press, New
  York, New York, USA, 2012, p. 185.
\newblock \href {https://doi.org/10.1145/2304576.2304601}
  {\path{doi:10.1145/2304576.2304601}}.

\bibitem{Menon2016}
H.~Menon, {Adaptive Load Balancing for HPC Applications}, Ph.D. thesis,
  University of Illinois Urbana-Champaign (2016).

\bibitem{Boulmier2017}
A.~Boulmier, I.~Banicescu, F.~Ciorba, N.~Abdennadher, {An autonomic approach
  for the selection of robust dynamic loop scheduling techniques}, in:
  Proceedings - 2017 IEEE 16th International Symposium on Parallel and
  Distributed Computing, ISPDC 2017, 2017.
\newblock \href {https://doi.org/10.1109/ISPDC.2017.9}
  {\path{doi:10.1109/ISPDC.2017.9}}.

\bibitem{MarquezClaudio2013AApplications}
{M{\'{a}}rquez Claudio}, {Eduardo C{\'{e}}sar}, {Joan Sorribes}, {A load
  balancing schema for agent-based spmd applications}, Proceedings of the
  International Conference on Parallel and Distributed Processing Techniques
  and Applications (PDPTA) (2013).

\bibitem{Procassini2004LoadCalculations}
R.~J. Procassini, M.~J. O'brien, J.~M. Taylor, {Load Balancing Of Parallel
  Monte Carlo Transport Calculations}, Tech. rep., International topical
  meeting on mathematics and computation, supercomputing, reactor physics and
  nuclear and biological applications, Avignon (France) (2004).

\bibitem{H.MenonandN.JainandG.ZhengandL.Kale2012}
{H. Menon and N. Jain and G. Zheng and L. Kal{\'{e}}}, {Automated Load
  Balancing Invocation Based on Application Characteristics}, in: 2012 IEEE
  International Conference on Cluster Computing, 2012, pp. 373--381.
\newblock \href {https://doi.org/10.1109/CLUSTER.2012.61}
  {\path{doi:10.1109/CLUSTER.2012.61}}.

\bibitem{Hart1968APaths}
P.~E. Hart, N.~J. Nilsson, B.~Raphael, {A formal basis for the heuristic
  determination of minimum cost paths}, IEEE transactions on Systems Science
  and Cybernetics 4~(2) (1968) 100--107.

\bibitem{Rodrigues2016StudyApplications}
F.~A. Rodrigues, \href{https://lume.ufrgs.br/handle/10183/149593}{{Study of
  load distribution measures for high-performance applications}}, Ph.D. thesis,
  Federal University of Rio Grande do Sul (2016).
\newline\urlprefix\url{https://lume.ufrgs.br/handle/10183/149593}

\bibitem{Offenhauser2017Load-balanceSystems}
P.~Offenh{\"{a}}user, {Load-balance strategies for CFD-codes on HPC systems},
  in: Proceedings of the 7th GACM Colloquium on Computational Mechanics for
  Young Scientists from Academia and Industry, OPUS, Stuttgart, Germany, 2017.
\newblock \href {https://doi.org/10.18419/opus-9334}
  {\path{doi:10.18419/opus-9334}}.

\bibitem{Lieber2018HighlyModeling}
M.~Lieber, W.~E. Nagel, {Highly scalable SFC-based dynamic load balancing and
  its application to atmospheric modeling}, Future Generation Computer Systems
  82 (2018) 575--590.
\newblock \href {https://doi.org/10.1016/j.future.2017.04.042}
  {\path{doi:10.1016/j.future.2017.04.042}}.

\bibitem{Ishiyama2012}
T.~Ishiyama, K.~Nitadori, J.~Makino, {4.45 Pflops astrophysical N-body
  simulation on K computer - The gravitational trillion-body problem}, in:
  International Conference for High Performance Computing, Networking, Storage
  and Analysis, SC, 2012.
\newblock \href {https://doi.org/10.1109/SC.2012.3}
  {\path{doi:10.1109/SC.2012.3}}.

\bibitem{Zhai2018}
K.~Zhai, T.~Banerjee, D.~Zwick, J.~Hackl, S.~Ranka, {Dynamic Load Balancing for
  Compressible Multiphase Turbulence}, in: Proceedings of the 2018
  International Conference on Supercomputing - ICS '18, ACM Press, New York,
  New York, USA, 2018, pp. 318--327.
\newblock \href {https://doi.org/10.1145/3205289.3205304}
  {\path{doi:10.1145/3205289.3205304}}.

\bibitem{DeRose2007DetectingSystems}
L.~DeRose, B.~Homer, D.~Johnson, {Detecting application load imbalance on high
  end massively parallel systems}, in: Lecture Notes in Computer Science
  (including subseries Lecture Notes in Artificial Intelligence and Lecture
  Notes in Bioinformatics), Vol. 4641 LNCS, Springer Verlag, 2007, pp.
  150--159.
\newblock \href {https://doi.org/10.1007/978-3-540-74466-5{\_}17}
  {\path{doi:10.1007/978-3-540-74466-5{\_}17}}.

\bibitem{XetqL/LBOPT:Parameters}
\href{https://github.com/xetqL/LBOPT}{{xetqL/LBOPT: Lightning fast code for
  computing load balancing scenario from application parameters}}.
\newline\urlprefix\url{https://github.com/xetqL/LBOPT}

\bibitem{XetqL/yalbb:Benchmark}
\href{https://github.com/xetqL/yalbb}{{xetqL/yalbb: Yet Another Load Balancing
  Benchmark}}.
\newline\urlprefix\url{https://github.com/xetqL/yalbb}

\bibitem{Tomczak2018SparseProcessors}
T.~Tomczak, R.~G. Szafran, {Sparse Geometries Handling in Lattice Boltzmann
  Method Implementation for Graphic Processors}, IEEE Transactions on Parallel
  and Distributed Systems 29~(8) (8 2018).
\newblock \href {https://doi.org/10.1109/TPDS.2018.2810237}
  {\path{doi:10.1109/TPDS.2018.2810237}}.

\bibitem{Top5002020}
\href{https://www.top500.org/lists/top500/2020/11/}{{Top500, November 2020}}.
\newline\urlprefix\url{https://www.top500.org/lists/top500/2020/11/}

\bibitem{Devine2002}
K.~Devine, E.~Boman, R.~Heaphy, B.~Hendrickson, C.~Vaughan, {Zoltan Data
  Management Services for Parallel Dynamic Applications}, Computing in Science
  and Engineering 4~(2) (2002) 90--97.

\bibitem{Latt2021Palabos:Solver}
J.~Latt, O.~Malaspinas, D.~Kontaxakis, A.~Parmigiani, D.~Lagrava, F.~Brogi,
  M.~B. Belgacem, Y.~Thorimbert, S.~Leclaire, S.~Li, F.~Marson, J.~Lemus,
  C.~Kotsalos, R.~Conradin, C.~Coreixas, R.~Petkantchin, F.~Raynaud, J.~Beny,
  B.~Chopard, {Palabos: Parallel Lattice Boltzmann Solver}, Computers {\&}
  Mathematics with Applications 81 (2021) 334--350.
\newblock \href {https://doi.org/https://doi.org/10.1016/j.camwa.2020.03.022}
  {\path{doi:https://doi.org/10.1016/j.camwa.2020.03.022}}.

\end{thebibliography}
\newpage
\appendix

\section{Optimal Scenario Finding Algorithm}
\begin{algorithm}
\KwIn{numIter: the number of iterations to compute, priorityQueue: a priority queue}
\SetAlgoLined

foundLb = [false$_0$, false$_1$, false$_2$,..., false$_{numIter}$]\;
\tcp{root node}
cNode = Node(iter=0, LB=true, cost=0.0, appState, lbState, prev=$\varnothing$)\; 
\While{cNode.iter $<$ numIter} {
    \If{cNode.LB} {
        foundLB[cNode.iter] = true\;
    }
    dontLBNode, doLBNode = cNode.getChildren()\;
    \If{not foundLB[doLBNode.iter]} {
        \tcp{Measurement of cost (i.e., time) with a theoretical model or a real application}
        doLBNode.computeCost()\;
        replaceOrInsertNode(priorityQueue, doLBNode)\;
    }
    dontLBNode.computeCost()\;
    insert(priorityQueue, dontLBNode)\;
    cNode = priorityQueue.pop()\;
}
\caption{Optimal Load Balancing Scenario Searching Algorithm}
\label{alg:get_optimum}
\end{algorithm}

\newpage

\section{Replace or Insert Algorithm}
\begin{algorithm}
\KwIn{priorityQueue: the priority queue, doLBNode: the load balancing node to insert or replace}
\SetAlgoLined
\For{node $\in$ priorityQueue} {
    \If{node.iter == doLBNode.iter  and node.LB == true} {
        \If{node.cost $>$ doLBNode.cost}{
            priorityQueue.remove(node)\;
            priorityQueue.insert(doLBNode)\;
        }
        return\;
    }
}
priorityQueue.insert(doLBNode)\;
\caption{replaceOrInsertNode(priorityQueue, doLBNode) : void}
\end{algorithm}
\label{alg:replaceinsert}
\end{document}